\documentclass{aa}  

\usepackage{graphicx}
\usepackage{txfonts}

\usepackage[T1]{fontenc}
\usepackage{ae,aecompl}
\usepackage[caption=false]{subfig}
\usepackage{graphicx}	
\usepackage{amsmath}	
\usepackage{amssymb}	
\usepackage{amsbsy}
\usepackage{color}
\usepackage{xcolor}
\usepackage{CJKutf8}
\usepackage{booktabs}
\usepackage{lipsum}
\usepackage{rotating}
\usepackage{hyperref}
\usepackage{xspace}

\newcommand{\dr}{{\mathrm{d}}}
\newcommand{\lgt}{\log_{10}}

\newcommand{\cco}{$\mathrm{CIB}\times \mathrm{KiDS}\,$}
\newcommand{\ccosfrd}{$\mathrm{CIB}\times \mathrm{KiDS}+{\rho_{\mathrm{SFR}}}\,$}
\newcommand{\planck}{{\it{Planck}}\,}

\newcommand{\euclid}{{\it{Euclid}}}

\newcommand{\healpix}{\textsc{healpix}\xspace}

\defcitealias{maniyar_simple_2021}{M21}

\begin{document}

   \title{Cosmic star formation history with tomographic cosmic infrared background{-}galaxy cross-correlation}

    \authorrunning{Yan et al.}

   \author{Ziang Yan 
          \inst{1,2}\thanks{E-mail:yanza21@astro.rub.de}
          \and Ludovic van Waerbeke\inst{2}
            \and Angus H. Wright \inst{1} 
            \and Maciej Bilicki\inst{3}
            \and Shiming Gu \inst{2}
            \and Hendrik Hildebrandt\inst{1}
            \and Abhishek S. Maniyar\inst{4}
            \and Tilman Tröster \inst{5}
          }

   \institute{Ruhr University Bochum, Faculty of Physics and Astronomy, Astronomical Institute (AIRUB), German Centre for Cosmological Lensing, 44780 Bochum, Germany
             \and
             Department of Physics and Astronomy, University of British Columbia, 6224 Agricultural Road, Vancouver, BC, V6T 1Z1, Canada
             \and
             Center for Theoretical Physics, Polish Academy of Sciences, al. Lotników 32/46, 02-668 Warsaw, Poland
             \and 
             Center for Cosmology and Particle Physics, Department of Physics, New York University, New York, NY 10003, USA
             \and 
             Institute for Astronomy, University of Edinburgh, Royal Observatory, Blackford Hill, Edinburgh, EH9 3HJ, UK
             }

   \date{}

 
\abstract{In this work we present a new method for probing the star formation history of the Universe, namely tomographic cross-correlation between the cosmic infrared background (CIB) and galaxy samples. The galaxy samples are from the Kilo-Degree Survey (KiDS), while the CIB maps are made from \planck\, sky maps at 353, 545, and 857 GHz. We measure the cross-correlation in harmonic space within $100<\ell<2000$ with a significance of 43$\sigma$. We model the cross-correlation with a halo model, which links CIB anisotropies to star formation rates (SFRs) and galaxy abundance. We assume that the SFR has a lognormal dependence on halo mass and that the galaxy abundance follows the halo occupation distribution (HOD) model. The cross-correlations give a best-fit maximum star formation efficiency of $\eta_{\mathrm{max}}= 0.41^{+0.09}_{-0.14}$ at a halo mass $\log_{10}(M_{\mathrm{peak}}/M_{\odot})= {12.14\pm 0.36}$. The derived star formation rate density (SFRD) is well constrained up to $z\sim 1.5$. The constraining power at high redshift is mainly limited by the KiDS survey depth. We also show that the constraint is robust to uncertainties in the estimated redshift distributions of the galaxy sample. A combination with external SFRD measurements from previous studies gives $\log_{10}(M_{\mathrm{peak}}/M_{\odot})=12.42^{+0.35}_{-0.19}$. This tightens the SFRD constraint up to $z=4$, yielding a peak SFRD of $0.09_{-0.004}^{+0.003}\,M_{\odot} \mathrm { year }^{-1} \mathrm{Mpc}^{-3}$ at $z=1.74^{+0.06}_{-0.02}$, corresponding to a lookback time of $10.05^{+0.12}_{-0.03}$ Gyr. Both constraints are consistent, and the derived SFRD agrees with previous studies and simulations. This validates the use of CIB tomography as an independent probe of the star formation history of the Universe. Additionally, we estimate the galaxy bias, $b$, of KiDS galaxies from the constrained HOD parameters and obtain an increasing bias from $b=1.1_{-0.31}^{+0.17}$ at $z=0$ to $b=1.96_{-0.64}^{+0.18}$ at $z=1.5$, which highlights the potential of this method as a probe of galaxy abundance. Finally, we provide a forecast for future galaxy surveys and conclude that, due to their considerable depth, future surveys will yield a much tighter constraint on the evolution of the SFRD.}

     \titlerunning{KiDS-CIB}
     
\maketitle
%
\section{Introduction}

Understanding the star formation activity in galaxies is central to our understanding of the evolution of galaxies in the Universe \citep[][]{1980FCPh....5..287T}. Moreover, the observed relationship between star formation and other physical processes implies that there exists complex interactions within galaxies between gases, stars, and central black holes (for example, through feedback from supernovae or supermassive black holes). Star formation activity can be described by the star formation rate density (SFRD), defined as the stellar mass generated per year per unit volume. By studying the SFRD of galaxies at different redshifts, we can understand the cosmic star formation history. In the local Universe, the star formation rate (SFR) can be explored via imaging the molecular gas in nearby galaxies \citep{2014sfrmc,2021phangs}. For distant galaxies, the SFRD is typically studied via multi-wavelength observations \citep{2013ruppioni, 2013magnelli,2016davies, 2016MNRAS.456.1999M}, which focus on the emission properties of galaxy populations over a wide range of wavelengths. By assuming the luminosity functions and the luminosity-stellar mass relations at different wavelengths, from ultraviolet (UV) to far-infrared, these works derive the SFR of galaxy populations from multi-wavelength luminosities. Previous multi-wavelength studies have reached a consensus (see \citealt{madau_cosmic_2014} for a review). From these studies, we are confident that star formation in the Universe started between  $6\lesssim z \lesssim 20$ and reached its peak at $z\sim 2$ (corresponding to a lookback time of $\sim 10.3$\,Gyr), at a rate approximately ten times higher than what is observed today. Due to a subsequent deficiency of available gas fuel, star formation activity has been steadily decreasing since $z\sim 2$.

Dust in stellar nurseries and in the interstellar medium of galaxies absorbs UV radiation produced by short-lived massive stars and re-radiates this energy as infrared (IR) emission \citep{kennicutt1998star, 2007ApJ...656..770S,2009sfrirg}. Therefore, extragalactic IR radiation can be used to study star formation throughout cosmic history. Moreover, one can also use the spectral energy distribution (SED) of IR galaxies to study the thermodynamics of interstellar dust in these galaxies \citep{bethermin_evolution_2015, bethermin_impact_2017}. However, dusty star-forming galaxies beyond the local Universe are typically highly blended, given the sensitivity and angular resolution of modern IR observatories, because they are both faint and numerous. {This makes it more difficult to study them individually in terms of their SFRD at higher redshift \citep[see, for example,][]{Nguyen_2010}} \footnote{Next-generation IR observatories, such as the James Webb Space Telescope, may make this direct probe more valuable, given their higher angular resolution.}.

Conversely, the projected all-sky IR emission, known as the cosmic infrared background \citep[CIB;][]{1967ApJ...147..868P}, encodes the cumulative emission of all dusty star-forming galaxies below $z\sim 6$. It is therefore a valuable tool that can be leveraged to investigate the SFRD. However, measurements of the CIB itself are complicated: imperfect removal of point sources and foreground Galactic emission can lead to bias in the measured CIB signals. Nevertheless, measurement of the CIB is more robust to the various selection effects and sample variance uncertainties that affect galaxy-based studies, and which depend highly on instrumental setup and observation strategies. The CIB was first detected by the Cosmic Background Explorer satellite \citep{1998cobecib} and was then accurately analysed by \textit{Spitzer} \citep{2006spitzercib} and \textit{Herschel} \citep{2010A&A...518L..30B}, {via large IR galaxy samples}, and by \planck\,\citep{2014planckxxx}, {via mapping of the CIB from its sky maps}. As with the cosmic microwave background (CMB), the key to {mapping} the CIB is accurately removing the foreground Galactic emission. {However, unlike the CMB, the peak frequency of the CIB is around $\sim 3000$ GHz, which is dominated by the thermal emission from Galactic dust.} Moreover, unlike the CMB and the thermal Sunyaev-Zel'dovich (tSZ) effect \citep{sunyaev1972observations}, the CIB has no unique frequency dependence. Therefore, the CIB is more difficult to extract from raw sky maps than the CMB or the tSZ effect and is generally restricted to high Galactic latitudes. There are multiple published CIB maps that are generated by various methods (differing primarily in the removal of the foreground). \citet{2014planckxxx} and \citet{lenz_large-scale_2019} remove Galactic emission by introducing a Galactic template from HI measurements, while \citet{2016planckcib} disentangles the CIB from Galactic dust by the generalised needlet internal composition \citep[GNILC;][]{Remazeilles_2011} method. The mean CIB emission measured from maps gives the mean energy emitted from star formation activities, while the anisotropies of the CIB trace the spatial distribution of star-forming galaxies and the underlying distribution of their host dark matter halos (given some galaxy bias). Therefore, analysing the CIB anisotropies via angular power spectra has been proposed as a new method to probe cosmic star formation activity \citep{2006spitzercib}. 

Angular power spectra have been widely used in cosmological studies. By cross-correlating different tracers, one can study cosmological parameters \citep{Kuntz_2015,singh2017cross}, the cosmic thermal history \citep{pandey2019constraints,koukoufilippas2020tomographic,chiang2020cosmic,2021yanz}, baryonic effects \citep{hojjati2017cross,troster2021joint}, the integrated Sachs-Wolfe (ISW) effect {\citep{Maniyar_2019, 2020hang}}, and more. Cosmic infrared background maps have been extensively used to study dusty star-forming galaxies via auto-correlations \citep{shang_improved_2012, 2014planckxxx, maniyar_star_2018} and cross-correlations with other large-scale structure tracers \citep{2020caoye, maniyar_simple_2021}. Clustering-based CIB cross-correlation has been used to study star formation in different types of galaxies; for example, \citet{Serra_2014} analyse luminous red galaxies (LRGs), \citet{2015MNRAS.449.4476W} analyse quasars, and \citet{2016ApJ...831...91C} analyse sub-millimetre galaxies (SMGs). The tracers used in these studies are either projected sky maps or galaxy samples with wide redshift ranges, leading to model parameters that describe the redshift dependence being highly degenerate. \citet{2014schmidt} and \citet{2018hall} cross-correlate the CIB with quasars at different redshifts, yielding an extensive measurement of the evolution of the CIB signal in quasars. However, these studies are restricted to active galaxies and therefore may miss contributions from the wider population of galaxies. 

This paper proposes a new clustering-based measurement that allows us to study the cosmic star formation history with the CIB: tomographic cross-correlation between the CIB and galaxy number density fluctuations. That is, cross-correlating the CIB with galaxy samples in different redshift ranges (so-called tomographic bins) to measure the evolution of the CIB over cosmic time. Compared with other large-scale structure tracers, galaxy number density fluctuations can be measured more directly. Firstly, galaxy redshifts can be determined directly via spectroscopy, although this process is expensive and must be restricted to particular samples of galaxies and/or small on-sky areas. Alternatively, wide-area photometric surveys provide galaxy samples that are larger and deeper than what can be observed with spectroscopy, and whose population redshift distribution can be calibrated to high accuracy with various algorithms (see \citealt{salvato2018flavours} for a review). Successful models have been proposed to describe galaxy number density fluctuations. On large scales, the galaxy distribution is proportional to the underlying mass fluctuation; on small scales, its non-linear behaviour can be modelled by a halo occupation distribution \citep[HOD;][]{Zheng_2005} model. With all these practical and theoretical feasibilities, galaxy density fluctuations have long been used to study various topics in large-scale structure, {including re-ionisation \citep{Lidz_2008}, cosmological parameters \citet{Kuntz_2015}, and the ISW effect \citep[][]{hang2020galaxy}.} In the near future,  the Canada-France Imaging Survey \citep[CFIS;][]{2017cfis}, the Rubin Observatory Legacy Survey of Space and Time \citep[LSST;][]{lsstsciencecollaboration2009lsst}, and the \textit{Euclid} \citep{laureijs2010euclid} mission will reach unprecedented sky coverage and depth, making galaxy number density fluctuation a `treasure chest' from which we will learn a lot about our Universe.

The CIB is generated from galaxies and so should correlate with galaxy distribution. Limited by the depth of current galaxy samples, CIB-galaxy cross-correlations are only sensitive to the CIB at low redshift, but this will improve with future galaxy surveys. In this study, we cross-correlate the galaxy catalogues provided by the Kilo-degree Survey \citep[KiDS;][]{de2013kilo} with CIB maps constructed at 353, 545, and 857 GHz to study the SFRD. The galaxy samples are divided into five tomographic bins extending to $z\sim 1.5$. Although the {measurements are} straightforward, modelling the CIB is more challenging than many other tracers. Firstly, SFRs and dust properties are different from galaxy to galaxy, and we do not have a clear understanding of both in a unified way. Previous studies take different models for the CIB: \citet{2014planckxxx} and \citet{shang_improved_2012} use a halo model by assuming a lognormal luminosity-to-halo mass ($L-M$) relation for the IR and a grey-body spectrum for extragalactic dust; \citet{maniyar_star_2018} and \citet{2020caoye} use the linear perturbation model with empirical radial kernel for the CIB; and \citet{maniyar_simple_2021} propose an HOD halo model for the CIB. In this work we use the \citet{maniyar_simple_2021} (\citetalias{maniyar_simple_2021} hereafter) model since it explicitly links the redshift dependence of the CIB with the SFR. 

This paper is structured as follows. In Sect.~\ref{sect:model} we describe the theoretical model we use for the cross-correlations. Sect.~\ref{sect:data} introduces the dataset that we are using. Section~\ref{sect:measurements} presents the method for measuring cross-correlations, as well as our estimation of the covariance matrix, likelihood, and systematics. Section \ref{sect:results} presents the results. Section~\ref{sect:discussions} discusses the results and summarises our conclusions. Throughout this study, we assume a flat $\Lambda$ cold dark matter cosmology with the fixed cosmological parameters from \citet{planckcosmo18} as our background cosmology: $ (h,\Omega_\mathrm{c} h^2,\Omega_\mathrm{b} h^2, \sigma_8,n_\mathrm{s}) = (0.676, 0.119, 0.022, 0.81, 0.967)$. 

\section{Models}
\label{sect:model}

\subsection{Angular cross-correlation}
\label{subsect:ccgeneral}

Both the galaxy and the CIB angular distributions are formalised as the line-of-sight projection of their 3D distributions. This subsection introduces the general theoretical framework of the angular cross-correlation between two projected cosmological fields. For an arbitrary cosmological field $u$, the projection of its 3D {fluctuations} (i.e. anisotropies) is written as

\begin{equation}
    \Delta_u(\hat{\theta})=\int\dr \chi W^{u}(\chi) \delta_u(\chi\hat{\theta}, \chi),
    \label{eq:fluctproj}
\end{equation}
where $\Delta_u(\hat{\theta})$ is the 2D projection in the angular direction $\hat{\theta}$, and $\delta_u(\chi\hat{\theta}, \chi)$ is the fluctuation of $u$ in 3D space at the coordinate $(\chi\hat{\theta}, \chi)$, where $\chi$ is the comoving distance. The kernel $W^{u}(\chi)$ describes the line-of-sight distribution of the field \footnote{The definition of the radial kernel can be quite arbitrary since, in practice, it can absorb any factors in the 3D field term that {{only}} depend on redshift. In the following subsections, we define the galaxy and the CIB kernels to emphasise the physical meaning of the 3D fields studied in this work.}

We measure the angular cross-correlation in harmonic space. In general, the angular cross-correlation between two projected fields, $u$ and $v$, at the scales $\ell \gtrsim 10$ are well estimated by the Limber approximation \citep{limber1953analysis, kaiser1992weak}: 

\begin{equation}
    C_{\ell}^{uv}  =  \int _0 ^{\chi_{\mathrm{H}}} \frac{\mathrm{d}\chi}{\chi^2}W^u(\chi)W^v(\chi)P_{uv}\left(k=\frac{\ell+1/2}{\chi}, z(\chi)\right), 
\end{equation}
where $P_{uv}(k, z)$ is the 3D cross-power spectrum of associated 3D fluctuating fields $u$ and $v$:

\begin{equation}
    \left\langle \delta_u(\boldsymbol{k})\delta_v(\boldsymbol{k}') \right\rangle = (2\pi)^3\delta(\boldsymbol{k}+\boldsymbol{k}')P_{uv}(k).
\label{eq:tracer_2d}
\end{equation}

\begin{figure}
    \centering
    \includegraphics[width=\columnwidth]{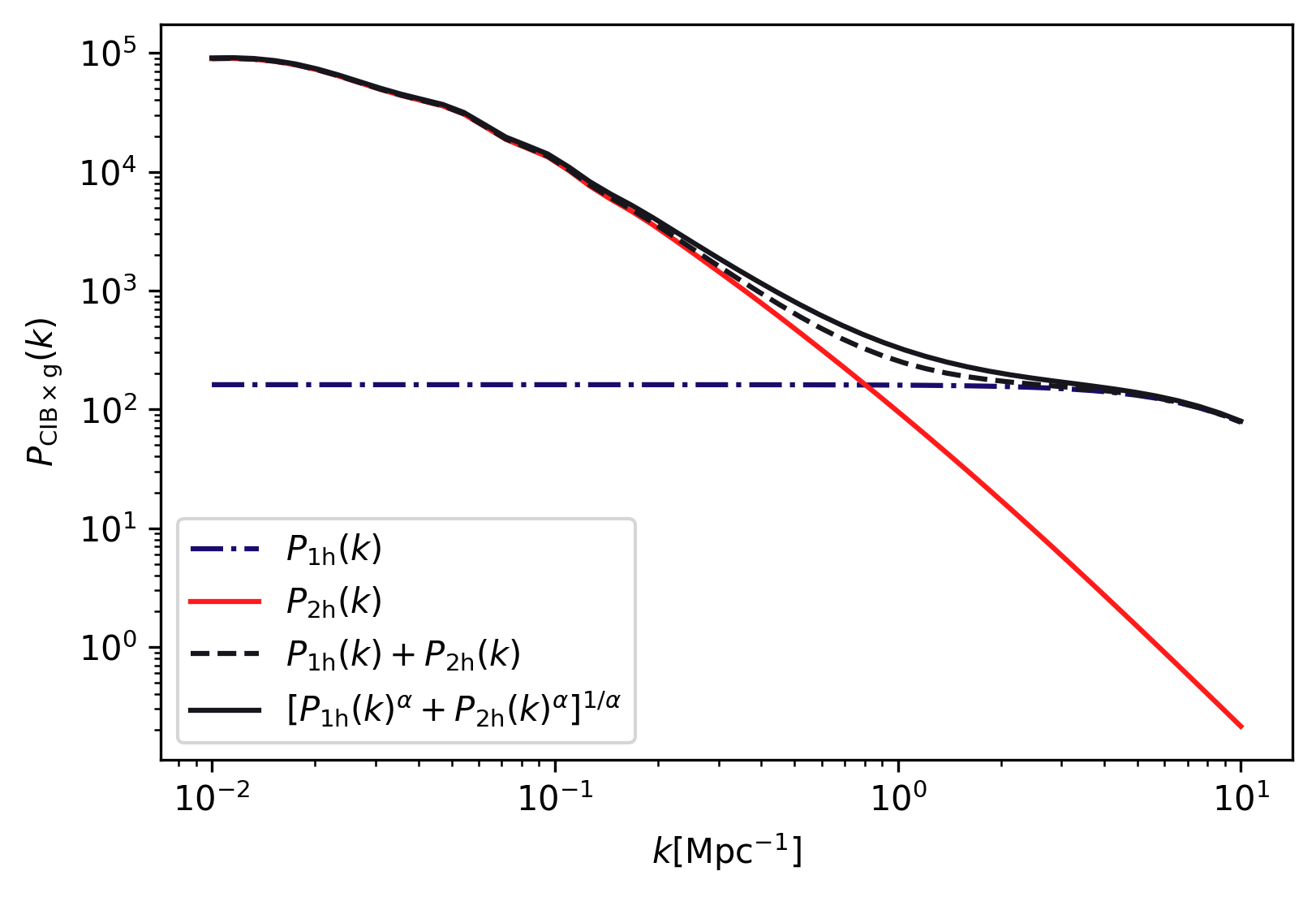}
    \caption{Halo model of the power spectrum of CIB-galaxy cross-correlation at $z=0$. The power spectrum in this plot only shows the spatial dependence of the correlation between the CIB and galaxy distribution, with all the irrelevant terms (redshift and frequency dependence) factored out, so the unit is arbitrary. The dash-dotted purple line and the solid red line are one- and two-halo terms, respectively; the dashed black line is the summation of one- and two-halo terms, and the solid black line is the smoothed power spectrum defined in Eq.~\eqref{eq:pk_smth}.}
    \label{fig:pk_hm}
\end{figure}

Generally, we can model a large-scale cosmological field as a biased-tracer of the underlying mass, mainly in the form of dark matter halos \citep{COORAY_2002, Seljak_2000}. In such a halo model, $P_{uv}(k)$ is divided into the two-halo term, which accounts for the correlation between different halos, and the one-halo term, which accounts for correlations within the same halo. \citet{Smith_2003} points out that simply adding the one- and two-halo up yields a total power spectrum that is lower than that from cosmological simulations in the transition regime ($k\sim 0.5 \,h\mathrm{Mpc}^{-1}$). \citet{mead2021hmcode2020} estimates that this difference can be up to a level of 40\%, so one needs to introduce a smoothing factor $\alpha$ to take this into account. The total power spectrum is then given by

\begin{equation}
    P_{uv}(k) = \left[P_{uv, \mathrm{1h}}(k)^{\alpha} + P_{uv, \mathrm{2h}}(k)^{\alpha}\right]^{1/\alpha}.
    \label{eq:pk_smth}
\end{equation}
The redshift dependence of $\alpha$ is given by the fitting formula in \citet{mead2021hmcode2020} \footnote{We note that the fitting formula originally applies to the matter power spectrum. There is no such formula for the CIB-galaxy power spectrum yet. This is an limitation of our formulation.}. In Fig. \ref{fig:pk_hm}, we plot the one- and two-halo terms (dash-dotted purple and red solid lines, respectively) of the CIB-galaxy cross-correlation power spectrum (to be introduced below), as well as their sum (the dashed black line) and the smoothed power spectrum (the solid black line). It is clear that the smoothing changes the power spectrum in the transition regime.

Both one- and two-halo terms are related to the profiles of $u$ and $v$ in Fourier space:

\begin{equation}
\begin{aligned}
P_{uv, \mathrm{1h}}(k) &=\int_0^{\infty} \dr M \frac{\dr n}{\dr M}\langle p_u(k | M) p_v(k | M)\rangle \\
P_{uv,\mathrm{2h}}(k) &=\langle b_u\rangle(k)\langle b_v\rangle(k) P^{\mathrm{lin}}(k) \\
\langle b_u\rangle(k) & \equiv \int_0^{\infty} \dr M \frac{\dr n}{\dr M} b_{\mathrm{h}}(M)\langle p_u(k | M)\rangle,
\end{aligned}
\label{eq:halomodel_pk}
\end{equation}
where the angled brackets $\langle \cdot \rangle$ describe the ensemble average of the quantity inside. {At a given redshift,} $P^{\mathrm{lin}}(k)$ is the linear power spectrum, $\dr n/\dr M$ is the halo mass function (number density of dark matter halos in each mass bin), $b_{\mathrm{h}}$ is the halo bias, and $p_u(k | M)$ is the profile of the tracer $u$ with mass $M$ in Fourier space:
\begin{equation}
p_u(k | M) \equiv 4 \pi \int_{0}^{\infty} \dr r r^{2} \frac{\sin (k r)}{k r} p_u(r | M),
\end{equation}
where $p_u(r | M)$ is the radial profile of $u$ in real space. In this work, we employ the halo mass function and halo bias given by \citet{Tinker_2008} and \citet{Tinker_2010}, respectively, in accordance with \citetalias{maniyar_simple_2021}.

\subsection{Galaxy number density fluctuations}
\label{subsect:gcib-cc}

The 2D projected galaxy number density fluctuation is measured as
\begin{equation}
    \Delta_{\mathrm{g}}(\hat{\theta})\equiv \frac{n(\hat{\theta})-\bar{n}}{\bar{n}},
\end{equation}
where $n(\hat{\theta})$ is the surface number density of galaxies {in} the direction $\hat{\theta}$ on sky, and $\bar{n}$ is the average surface number density. Given the redshift distribution of a galaxy sample $\Phi_{\mathrm{g}}(z)$ (determined by the true line-of-sight galaxy distribution and any survey selection functions)\footnote{In the literature, galaxy redshift distributions are typically denoted as $n(z)$. In this paper, in order to prevent confusion with galaxy number densities, we instead use $\Phi_{\mathrm{g}}(z)$ to denote survey-specified galaxy redshift distributions.}, the projected galaxy density fluctuation is given by

\begin{equation}
\Delta_{\mathrm{g}}(\hat{\boldsymbol{\theta}})=
\int_0 ^{\chi_{\mathrm{H}}} \mathrm{d} \chi \frac{H(z)}{c}\Phi_{\mathrm{g}}(z(\chi)) \delta_{\mathrm{g}}(\chi(z) \hat{\boldsymbol{\theta}}, \chi)
,
\label{eq:gal_2d}
\end{equation}
where $\delta_{\mathrm{g}}(\chi(z) \hat{\boldsymbol{\theta}}, \chi)$ is the 3D galaxy density fluctuation. The radial kernel for galaxy number density fluctuation is then
\begin{equation}
W^{\mathrm{g}}(\chi) = \frac{H(\chi)}{c} {\Phi_{\mathrm{g}}[z(\chi)]}
\label{eq:wg}
.\end{equation}

The galaxy density fluctuation in a halo with mass $M$ can be described by its number density profile $p_{\mathrm{g}}(r| M)$, as 
\begin{equation}
\begin{aligned}
        \delta_{\mathrm{g}}(r | M)&=\frac{1}{\bar{n}_{\mathrm{g}}(z)} p_{\mathrm{g}}(r| M)\\ &=\frac{1}{\bar{n}_{\mathrm{g}}(z)}\left[ N_{\mathrm{c}}(M)\delta^{\mathrm{3D}}(\boldsymbol{r})+ N_{\mathrm{s}}(M) p_{\mathrm{s}}(r | M)\right],
\end{aligned}
\label{eq:delta_g}
\end{equation}
where $\delta^{\mathrm{3D}}$ is the 3D Dirac delta function, $N_{\mathrm{c}}(M)$ and $N_{\mathrm{s}}(M)$ are the number of central galaxy and satellite galaxies as a function of the halo mass ($M$), {respectively}, and $p_{\mathrm{s}}(r | M)$ is the number density profile of the satellite galaxies. Its Fourier transform will be given in Eq.~\eqref{eq:p_s}.  $\bar{n}_{\mathrm{g}}(z)$ is the mean galaxy number density at redshift $z$, which is given by

\begin{equation}
    \bar{n}_{\mathrm{g}} = \int \dr M \frac{\dr n}{\dr M}(N_{\mathrm{c}}(M)+N_{\mathrm{s}}(M)).
\end{equation}
Though we cannot say anything about galaxy counts for individual halos, their ensemble averages can be estimated via the HOD model \citep{Zheng_2005,Peacock_2000}:

\begin{equation}
\begin{aligned}
     \langle N_{\mathrm{c}}(M) \rangle &= \frac{1}{2}\left[1+\operatorname{erf}\left(\frac{\ln \left(M / M_{\min }\right)}{\sigma_{\ln M}}\right)\right] \\
     \langle N_{\mathrm{s}}(M) \rangle &= N_{\mathrm{c}}(M) \Theta\left(M-M_{0}\right)\left(\frac{M-M_{0}}{M_{1}}\right)^{\alpha_{\mathrm{s}}},
\end{aligned}
\label{eq:ngal_hod}
\end{equation}
{where $M_{\min }$ is the mass scale at which half of all halos host a galaxy, $\sigma_{\ln M}$ denotes the transition smoothing scale, $M_1$ is a typical halo mass that consists of one satellite galaxy, $M_0$ is the threshold halo mass required to form satellite galaxies, and $\alpha_{\rm s}$ is the power law slope of the satellite galaxy occupation distribution. $\Theta$ is the Heaviside step function.}

In Fourier space, the galaxy number profile is given by
\begin{equation}
    \delta_{\mathrm{g}}(k| M)=\frac{1}{\bar{n}_{\mathrm{g}}(z)} p_{\mathrm{g}}(k| M) =\frac{1}{\bar{n}_{\mathrm{g}}(z)}\left[ N_{\mathrm{c}}(M)+ N_{\mathrm{s}}(M) p_{\mathrm{s}}(k | M)\right],
\label{eq:gprof_hod}
\end{equation}
where the dimensionless profile of satellite galaxies $p_{\mathrm{s}}(k | M)$ is generally taken as the Navarro-Frenk-White (NFW) profile (\citealt{van_den_Bosch_2013}; \citealt{Navarro_1996}):
\begin{equation}\label{eq:p_s}
\begin{aligned}
p_{\mathrm{s}}(k | M) = p_{\mathrm{NFW}}(k \mid M)=& {\left[\ln \left(1+c\right)-\frac{c}{\left(1+c\right)}\right]^{-1} } \\
& \times {\left[\cos (q)\left(\operatorname{Ci}\left(\left(1+c\right) q\right)-\operatorname{Ci}(q)\right)\right.} \\
&+\sin (q)\left(\operatorname{Si}\left(\left(1+c\right) q\right)-\operatorname{Si}(q)\right) \\
&\left.-\sin \left(cq\right) /\left(1+cq\right)\right]
\end{aligned}
,\end{equation}
where $q\equiv kr_{200}(M)/c(M)$, $c$ is the concentration factor, and the functions \{Ci, Si\} are the standard cosine and sine integrals, respectively\footnote{{The cosine and sine integrals are defined as follows: \[\mathrm{Ci}(x)\equiv \int_x^{\infty}\frac{\cos t}{t}\mathrm{d}t,\]\[\mathrm{Si}(x)\equiv \int_0^{x}\frac{\sin t}{t}\mathrm{d}t.\]}}. The concentration-mass relation in this work is given by \citet{Duffy_2008}. Here $r_{200}$ is the radius that encloses a region where the average density exceeds 200 times the critical density of the Universe. We take the total mass within $r_{200}$ as the proxy of halo mass because in general $r_{200}$ is close to the virial radius of a halo \citep[][]{1998tx19.confE.533O}.\footnote{In the literature, this mass is typically denoted as $M_{200}$, but we omit the subscript here.}

The HOD parameters in Eq.~\eqref{eq:ngal_hod} depend on redshift \citep[][]{2012cfht}. In this work, we fix $\sigma_{\ln M}=0.4$ and $\alpha_\mathrm{s}=1$, consistent with simulations \citep{Zheng_2005} {and previous observational constraints \citep[][]{2012cfht, Ishikawa_2020}}, and adopt a simple relation for $\{M_0, M_1, M_{\mathrm{min}}\}$ with respect to redshift. For example, we model $M_0$ as in \citet{2013ApJ...770...57B}:

\begin{equation}
    \log_{10}M_0(a) = \log_{10}M_{0,0} + \log_{10}M_{0,p}(a-1),
    \label{eq:mp}
 \end{equation}
where $a$ is the scale factor, $\log_{10}M_{0,0}$ is the value at $z=0$,  while $\log_{10}M_{0,p}$ gives the `rate' of evolution\footnote{In this paper, all the  logarithmic masses are assumed to be in the unit of $M_{\odot}$.}. Therefore, in total we constrain six HOD parameters: $\{M_{0,0}, M_{1,0}, M_{{\mathrm{min}},0}, M_{0,p}, M_{1,p}, M_{{\mathrm{min}},p}\}$. In practice, we find that the resolution of the CIB map is sufficiently low that this simple formalism fits the data well (Sect.~\ref{sect:results}).

\subsection{Halo model for CIB-galaxy cross-correlation}

The intensity of the CIB (in $\mathrm{Jy/sr}$) is the line-of-sight integral of the comoving emissivity, $j_{\nu}$:
\begin{equation}
    I_{\nu}(\boldsymbol{\theta}) = \int \mathrm{d}\chi a j_{\nu}(\chi, \boldsymbol{\theta}).
    \label{eq:cibintens}
\end{equation}
Comparing with Eq.~\eqref{eq:fluctproj}, one can define the radial kernel for the CIB to be

\begin{equation}
    W^{\mathrm{CIB}}(\chi) = a(\chi) = \frac{1}{1+z(\chi)},
\end{equation}
which is independent of frequency. Thus, the emissivity $j_{\nu}$ is the `$\delta_u$' for the CIB, which is related to the underlying galaxy population as

\begin{equation}
\begin{aligned}
j_{\nu}(z) &= \int \dr L \frac{\dr n}{\dr L}\frac{L_{(1+z)\nu}(z)}{4\pi} \\
&= \int \dr M \frac{\dr n}{\dr M}\frac{L_{(1+z)\nu}(M, z)}{4\pi},
\end{aligned}
\label{eq:j_l}  
\end{equation}
where $L_{\mathrm{\nu}}(z)$ is the IR luminosity and ${\dr n}/{\dr L}$ is the IR luminosity function. The second equation assumes that galaxy luminosity is also a function of the mass of the host dark matter halo. Furthermore, like galaxies, the model of the IR luminosity can also be divided into contributions from central and satellite galaxies \citep[][]{shang_improved_2012, 2014planckxxx}. We introduce the IR luminous intensity (i.e. the power emitted per steradian):

\begin{equation}
    f_{\nu,\mathrm{c/s}}(M, z) \equiv \frac{L_{(1+z)\nu, \mathrm{c/s}}(M, z)}{4\pi},
    \label{eq:f_l}
\end{equation}
where the subscripts `$\mathrm{c/s}$' denote the central and satellite components, respectively. The profile of the CIB in Fourier space is formulated as\begin{equation}
    f_{\nu}(k| M) = f_{\nu, \mathrm{c}}(M) + f_{\nu, \mathrm{s}}(M)p_{\mathrm{NFW}}(k | M).
    \label{eq:cibfcs}
\end{equation}
Comparing with Eq.~\eqref{eq:delta_g}, one recognises that the quantity $f_{\nu,\mathrm{c/s}}(M)$ is directly analogous to $N_{\mathrm{c/s}}(M)$, and $f_{\nu}(k| M)$ is the profile term $p_u(k| M)$ in {Eq.~}\eqref{eq:halomodel_pk} for CIB anisotropies. Following the standard practice of \citet{van_den_Bosch_2013}, we give the cross-correlation between the Fourier profile of galaxies and the CIB that is needed for calculating the one-halo term:

\begin{equation}
\begin{aligned}
\left\langle p_{g}(k| M) f_{\nu}(k| M)\right\rangle & =  \langle N_{s}(M) \rangle\langle f_{\nu, \mathrm{s}}(M)\rangle p^{2}(k| M)  \\ 
  & +  \langle N_{\mathrm{c}}(M)\rangle \langle f_{\nu, \mathrm{s}}(M)\rangle p(k| M) \\
  &+ \langle N_{\mathrm{s}}(M)\rangle \langle f_{\nu, \mathrm{\mathrm{c}}}(M)\rangle  p(k| M).
\end{aligned}
\label{eq:pj_1h}
\end{equation}
We discuss how to model $f_{\nu,\mathrm{c/s}}$ in Sect. \ref{subsec:jsfr}.

\subsection{CIB emissivity and star formation rate}
\label{subsec:jsfr}
Considering the origin of the CIB, $j_{\nu}$ should depend on the dust properties (temperature, absorption, etc.) of star-forming galaxies, and on their SFR. In addition, the CIB also directly traces the spatial distribution of galaxies and their host halos. We take the halo model for the CIB from \citetalias{maniyar_simple_2021}. The {observed} mean CIB emissivity at redshift $z$ is given by

\begin{equation}
    j_{\nu}(z) = \frac{{\rho_{\mathrm{SFR}}(z)}(1+z)S_{\mathrm{eff}}\left[(1+z)\nu, z\right]\chi^2}{K},
    \label{eq:jsfrd}
\end{equation}
where ${\rho_{\mathrm{SFR}}(z)}$ is the SFRD, defined as the stellar mass formed per year per unit volume (in $M_{\odot}\mathrm{yr}^{-1}\mathrm{Mpc}^{-3}$), and $S_{\mathrm{eff}}(\nu, z)$ is the effective SED of IR emission from galaxies at the given the {rest-frame} frequency $\nu$ and redshift $z$. {The latter term is defined as the mean flux density received from a source with $L_{\mathrm{IR}}=1L_{\odot}$, so it has a unit of $\mathrm{Jy}/L_{\odot}$.} We note that we change to the rest frame frequency by multiplying the observed frequency $\nu$ by $(1+z)$. The \citet[][]{kennicutt1998star} constant $K$ is defined as the ratio between SFRD and IR luminosity density. In the wavelength range of $8-1000\,\mu\mathrm{m}$, it has a value of $K=1\times 10^{-10}M_{\odot}\mathrm{yr}^{-1}L_{\odot}^{-1}$ {assuming a Chabrier initial mass function \citep[IMF;][]{2003PASP..115..763C}}. The derivation of this formula can be found in Appendix B of \citet{2014planckxxx}. 

The SFRD is given by

\begin{equation}
\rho_{\mathrm{SFR}}(z) =\int \mathrm{d}M \frac{\dr n}{\dr M} \mathrm{SFR}_{\mathrm{tot}}(M,z),
\label{eq:sfrd_z}
\end{equation}
where $\mathrm{SFR}_{\mathrm{tot}}(M,z)$ denotes the total SFR of the galaxies in a halo with mass $M$ at redshift $z$. 

As is shown in Eq.~\eqref{eq:cibfcs}, $f_{\nu}(M,z)$ can also be divided into components from the central galaxy and satellite galaxies living in sub-halos. Following \citet{shang_improved_2012, maniyar_star_2018, maniyar_simple_2021}, we assume that the central galaxy and satellite galaxies share the same effective SED. In the literature, the SED in {Eq.~}\eqref{eq:jsfrd} are given with different methods: \citet{shang_improved_2012} parametrises the effective SED with a grey-body spectrum,  while \citet{maniyar_star_2018, maniyar_simple_2021} use fixed effective SED from previous studies. In this work, we follow \citetalias{maniyar_simple_2021} and take the SED calculated with the method given by \citet{bethermin_redshift_2013}; that is, we assume the mean SED of each type of galaxies (main sequence, starburst), and weigh their contribution to the whole population in construction of the effective SED. The SED templates and weights are given by \citet{bethermin_evolution_2015, bethermin_impact_2017}. Therefore, central and satellite components differ only in SFR, and the total SFR in Eq.~\eqref{eq:sfrd_z} is given by $\mathrm{SFR}_{\mathrm{tot}}=\mathrm{SFR}_{\mathrm{c}}+\mathrm{SFR}_{\mathrm{s}}$. Combining Eqs.~\eqref{eq:j_l},\eqref{eq:f_l}, \eqref{eq:jsfrd}, and \eqref{eq:sfrd_z}, we recognise that

\begin{equation}
    f_{\nu,\mathrm{c/s}}(M,z) = \frac{{{\mathrm{SFR}_{\mathrm{c/s}}}(M,z)}(1+z)S_{\mathrm{eff}}\left[(1+z)\nu, z\right]\chi^2}{K}.
\end{equation}

The final piece of the puzzle for our model is in defining $\mathrm{SFR}_{\mathrm{c/s}}$. Following \citet{bethermin_redshift_2013}, the SFR is given by the baryon accretion rate (BAR, measured in solar masses per year: ${M_{\odot} \mathrm{yr}^{-1}}$) multiplied by the star formation efficiency $\eta$. That is,\begin{equation}
    \mathrm{SFR}(M, z)=\eta(M,z)\times \mathrm{BAR}(M, z).
\label{eq:sfr}
\end{equation}
{For a given halo mass $M$ at redshift $z$, the BAR is the mean mass growth rate (MGR; also measured in ${M_{\odot} \mathrm{yr}^{-1}}$) of the halo multiplied by the baryon-matter ratio:
\begin{equation}
\begin{aligned}
      \mathrm{BAR}(M, z) &=  \mathrm{MGR}(M, z) \\
&\times \frac{\Omega_\mathrm{b}}{\Omega_m}.
\end{aligned}
\end{equation}
The MGR is given by the fits of \citep{fakhouri2010merger}

\begin{equation}
\begin{aligned}
\mathrm{MGR}(M, z) = & 46.1 \left(\frac{M}{10^{12} M_{\odot}}\right)^{1.1} \times \\
& (1+1.11 z) \sqrt{\Omega_{\mathrm{m}}(1+z)^{3}+\Omega_{\Lambda}}
\end{aligned}
,\end{equation}
where $\Omega_{\mathrm{m}}$, $\Omega_{\mathrm{b}}$, and $\Omega_{\Lambda}$ are the density parameters of total mass, baryons, and dark energy of the Universe, respectively.}

The {star formation} efficiency is {parameterised} as a lognormal function of the halo mass, $M$:

\begin{equation}
\eta(M,z)=\eta_{\mathrm{max}} \exp{\left[-\frac{\left(\ln M-\ln M_{\mathrm{peak}}\right)^{2}}{2 \sigma_{M}(z)^{2}}\right]},
\label{eq:eta}
\end{equation}
where $M_{\mathrm{peak}}$ represents the mass of a halo with the highest star formation efficiency $\eta_{\mathrm{max}}$. {An analysis of average SFRs and histories in galaxies from $z=0$ to $z = 8$ shows that $M_{\mathrm{peak}}$ ought to be constant over cosmic time, at a value of  $M_{\mathrm{peak}}\sim 10^{12}M_{\odot}$ \citet{2013ApJ...770...57B}. Therefore, in our model, we assume it to be a constant.}  And $\sigma_{M}(z)$ is the variance of the lognormal, which represents the range of masses over which the star formation is efficient. Also, following \citetalias{maniyar_simple_2021}, this parameter depends both on redshift and halo mass:

\begin{equation}
    \sigma_M(z) =
  \begin{cases}
    \sigma_{M, 0} & \text{if $M<M_{\mathrm{peak}}$} \\
    \sigma_{M, 0}-\tau\times\max \{0, z_c-z\} & \text{if $M\geq M_{\mathrm{peak}}$} 
  \end{cases}
  \label{eq:sigma_m}
,\end{equation}
where $z_c$ is the redshift above which the mass window for {star formation} starts to evolve, with a `rate' described by a free parameter $\tau$. In this work, we fix $z_c=1.5$, as in \citetalias{maniyar_simple_2021}, because our sample of KiDS galaxies is unable to probe beyond this redshift (see Sect.~\ref{sect:data} and Fig.~\ref{fig:dndz}).

For the central galaxy, the SFR is calculated with Eq.~\eqref{eq:sfr}, where $M$ describes the mass of the host halo, multiplied by the mean number of central galaxies $\langle N_{\mathrm{c}}\rangle$ as given by Eq.~\eqref{eq:ngal_hod}:

\begin{equation}
    \mathrm{SFR}_{\mathrm{c}}(M) = \langle N_{\mathrm{c}}(M)\rangle\times\mathrm{SFR}(M)
.\end{equation}
\label{eq:sfr_c}

For satellite galaxies, the SFR depend on the masses of sub-halos in which  they are located \citep{bethermin_redshift_2013}:

\begin{equation}
    \mathrm{SFR}_{\mathrm{s}}(m| M) = \mathrm{min}\left\{\mathrm{SFR}(m), m/M\times\mathrm{SFR}(M)\right\},
\end{equation}
where $m$ is the sub-halo mass, and  $\mathrm{SFR}$ is the general SFR defined by Eq.~\eqref{eq:sfr}. The mean SFR for sub-halos in a host halo with mass $M$ is then 

\begin{equation}
    \mathrm{SFR_{s}}(M) = \int \mathrm{d}\ln{m} \frac{\dr N_{\mathrm{sub}}}{\dr \ln{m}}\mathrm{SFR}_{\mathrm{s}}(m|M),
\end{equation}
where ${\dr N_{\mathrm{sub}}}/{\dr \ln{m}}$ is the sub-halo mass function. We take the formulation given by \citet{2010tinkershmf}. Once we get the SFR for both the central and the sub-halos, we can add them together and calculate the luminous intensity $f_{\nu}$ of a halo with Eq.~\eqref{eq:jsfrd}, and then calculate the angular power spectra with the halo model and Limber approximation as discussed in {Sects.~}\ref{subsect:ccgeneral} and \ref{subsect:gcib-cc}. 

{There are a couple of simplifying assumptions in our model. First of all, we assume that the IR radiation from a galaxy is entirely the thermal radiation from dust{, which is generated by star formation activity.}. However, part of the IR radiation may be generated from non-dust radiation, including CO(3-2), free-free scattering, or synchrotron radiation \citep{Galametz_2014}. We also assume that central and satellite galaxies have the same dust SED, which might be entirely accurate. In addition, we neglect the difference in quenching in central and satellite galaxies \citep[][]{2015MNRAS.449.4476W}. However, the IR radiation is dominated by central galaxies, so the differences between central and satellite galaxies will not significantly affect our conclusion. In any case, these limitations need further investigation by future studies.}

We note, though, that the measured power spectrum will also contain shot-noise resulting from the auto-correlated Poisson sampling noise. Therefore, the model for the total CIB-galaxy cross-correlation {$C_{\ell}^{\nu\mathrm{g}, \mathrm{tot}}$} is

\begin{equation}
    C_{\ell}^{\nu\mathrm{g}, \mathrm{tot}} = C_{\ell}^{\nu\mathrm{g}, \mathrm{hm}} + S^{\nu\mathrm{g}},
    \label{eq:cltot}
\end{equation}
where $C_{\ell}^{\nu\mathrm{g}, \mathrm{hm}}$ is the cross-correlation predicted by the halo model, and $S^{\nu\mathrm{g}}$ is the scale-independent shot noise. Shot-noise is generally not negligible in galaxy cross-correlations, especially on small scales. {There are analytical models to predict shot noise \citep{2014planckxxx, 2018shot} but they depends on various assumptions, including the CIB flux cut, galaxy colours, galaxy physical evolution \citep{2012bethermin}, and so on. Each of these assumptions carries with it additional uncertainties. Therefore, in this work, {instead of modelling the shot noise for different pairs of cross-correlations}, we simply opt to keep their amplitudes as free parameters in our model \footnote{{In this study, we ignore the variable depth in different patches in the KiDS footprint. This variance alters the galaxy number in different patches in the KiDS footprint, which might cause spatial dependence of the shot-noise power spectra.}}. In practice, we set $\log_{10}{S^{\nu\mathrm{g}}}$ to be free, where $S^{\nu\mathrm{g}}$ is in the unit of $\mathrm{MJy/sr}$.}

With the SFR and SED models introduced above, the redshift distribution of the CIB intensity $\frac{\dr I_{\nu}}{\dr z}$ can be calculated with Eq.~\eqref{eq:cibintens}. The redshift distributions of the CIB intensity at 353, 545, and 857 GHz are shown as dotted lines in Fig. \ref{fig:dndz}. It is clear that CIB emission increases with frequency (in the frequency range we explore here). The peak redshift of CIB emission is $z\sim 1.5$, which is {close to the redshift limit of our galaxy sample}.

In addition, once we have fixed the model parameters in $\mathrm{SFR}$ with our measurements, we can calculate $\rho_{\mathrm{SFR}}$ by adding up SFR of central and satellite galaxies and employing Eq.~\eqref{eq:sfrd_z}: 
\begin{equation}
    \begin{split}
\rho_{\mathrm{SFR}}(z) &=\int \mathrm{d}M \frac{\dr n}{\dr M}\mathrm{SFR}_{\mathrm{tot}}(M,z)\\
&=\int \mathrm{d}M \frac{\dr n}{\dr M} \left[\mathrm{SFR}_{\mathrm{c}}(M, z)+\mathrm{SFR}_{\mathrm{s}}(M, z)\right] \\
&= \int \mathrm{d}M \frac{\dr n}{\dr M} \left[\vphantom{\int \mathrm{d}\ln{m} \frac{\dr N_{\mathrm{sub}}}{\dr \ln{m}}(m|M)\mathrm{SFR}_{\mathrm{s}}(m|M, z)}\langle N_{\mathrm{c}}(M)\rangle\mathrm{SFR}(M, z) \right. \\
&+ \left. \int \mathrm{d}\ln{m} \frac{\dr N_{\mathrm{sub}}}{\dr \ln{m}}(m|M)\mathrm{SFR}_{\mathrm{s}}(m|M, z)\right].
\end{split}
\end{equation}
The primary goal of this work is to constrain this $\rho_{\mathrm{SFR}}(z)$ parameter with CIB-galaxy cross-correlation.

\section{Data}
\label{sect:data}
\subsection{KiDS data}
\begin{figure*}
    \centering
    \includegraphics[width=0.8\textwidth]{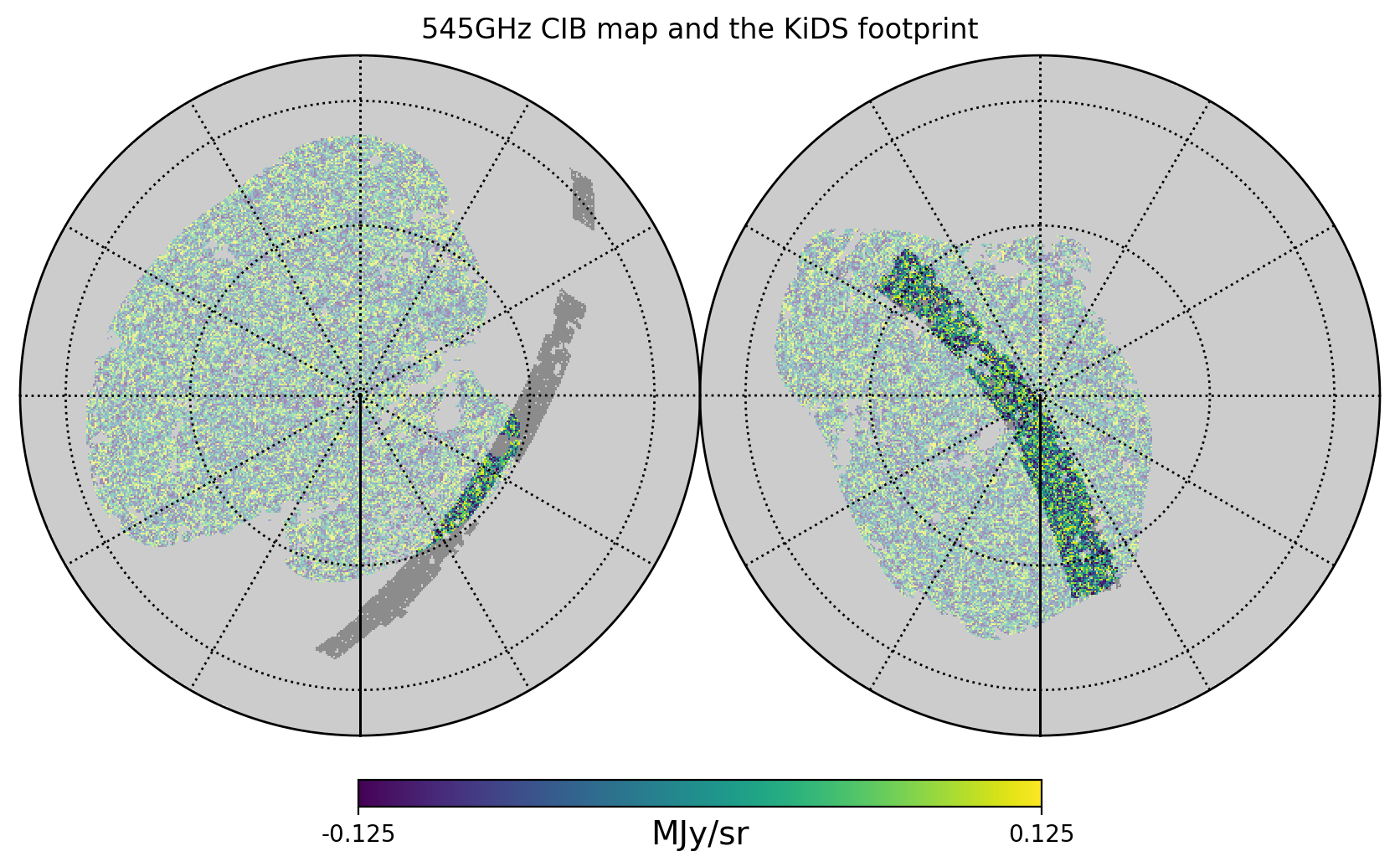}
    \caption{CIB map at 545 GHz for the Galactic north pole (left) and south pole (right). The transparent regions are the KiDS footprint.}
    \label{fig:masks}
\end{figure*}

We used the lensing catalogue provided by the fourth data release (DR4) of KiDS \citep{Kuijken_2019} as our galaxy sample. KiDS is a wide-field imaging survey that measures the positions and shapes of galaxies using the VLT Survey Telescope (VST) at the European Southern Observatory (ESO). Both the telescope and the survey were primarily designed for weak-lensing applications. The footprint of the KiDS DR4 (and its corresponding galaxy sample, called the `KiDS-1000' sample) is divided into a northern and southern patch, with total coverage of 1006 $\mathrm{deg}^2$ of the sky (corresponding to a fractional area of the full sky of $f_{\mathrm{sky}}=2.2\%$.) The footprint is shown as the transparent patches in Fig. \ref{fig:masks}. High-quality optical images are produced with VST-OmegaCAM, and these data are then combined with imaging from the VISTA Kilo-degree INfrared Galaxy (VIKING) survey (\citealt{2013Msngr.154...32E}), allowing all sources in KiDS-1000 to be photometrically measured in nine optical and near-IR bands: $ugriZYJHK_{\mathrm{s}}$ \citep[][]{2019A&A...632A..34W}. {The KiDS-1000 sample selects galaxies with photometric redshift estimates $0.1<z_\mathrm{B}\leq 1.2$.}  Although the sky coverage of KiDS is relatively small comparing to some galaxy surveys \citep[such as the Dark Energy Survey;][]{abbott2016dark}, galaxy photometric redshift estimation and redshift distribution calibration (especially at high redshift) is more reliable in KiDS thanks to the complimentary near-IR information from VIKING (which was co-designed with KiDS to reach complimentary depths in the near-IR bands). Each galaxy in the KiDS-1000 sample has ellipticities measured with the \emph{Lens}fit algorithm \citep{miller2013bayesian}, which allows exceptional control for systematic effects such as stellar contamination \citep[][]{giblin2020kids1000}. The KiDS-1000 sample is then further down-selected during the production of high-accuracy calibrated redshift distributions \citep{Wright_2020,hildebr2020kids1000} to produce the KiDS-1000 `gold' sample \footnote{{The gold selection removes about 20\% of the source galaxies, which enhances the shot noise. An ongoing analysis shows that the spatial dependence of this selection is negligible. }}.  We used the gold sample for this work as the redshift distributions are most accurately calibrated for these galaxies. We present the information of the galaxy sample that we use in Table \ref{table:tomoinfo}. This resulting galaxy sample covers redshifts $z\lesssim 1.5$, and is therefore a suitable dataset to trace the history of different components of the {large-scale structure} into the intermediate-redshift Universe.

{Although the KiDS survey provides high-accuracy shape measurements for galaxies, we do not use them in this analysis. As is argued in \citet{2021yanz}, galaxy number density fluctuations are relatively easy to measure (compared to galaxy shapes) and are immune to the systematic effects inherent to the shape measurement process (including shape measurement accuracy, response to shear, shape calibration error, intrinsic alignment, etc). Moreover, the CIB is generated directly from galaxies, so we expect strong CIB-galaxy correlation signals, which reveal the star formation activity in these galaxies. Therefore, we focus on CIB-galaxy cross-correlation in this work, allowing us to ignore shape information. However, we note that \citet{troster2021joint} has made a significant detection of shear-CIB cross-correlation with the 545 GHz \planck\, CIB map, which can help us understand the connection between halos and IR emissions. We leave such an investigation for future work.}

\begin{figure}
    \centering
    \includegraphics[width=\columnwidth]{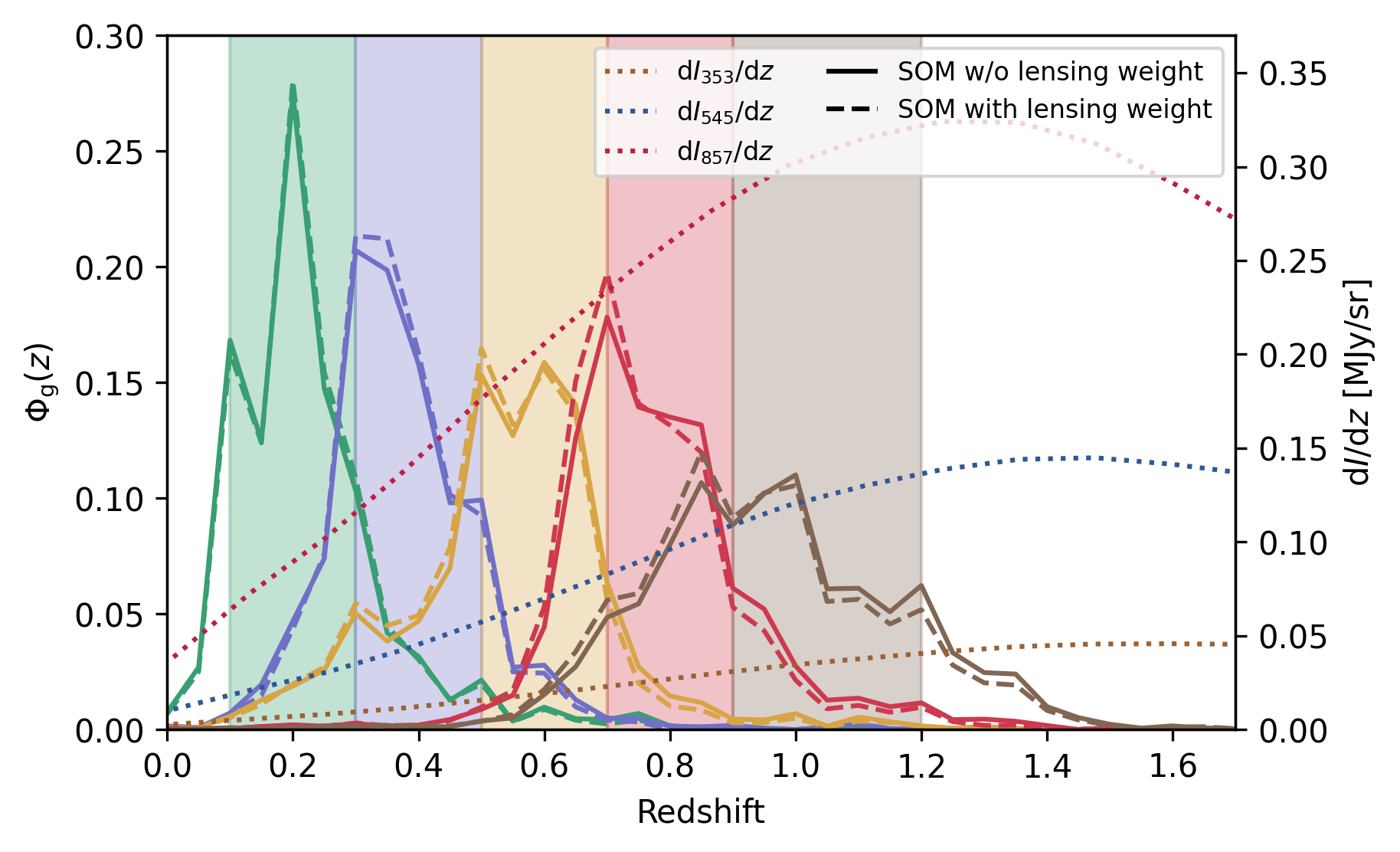}
    \caption{Redshift distributions of the KiDS-1000 gold galaxy sample (solid and dashed lines) and CIB emissions (dotted lines). Solid lines are the redshift distribution of the KiDS galaxies calibrated by the SOM without specifying lensing weight and are the redshift distributions we use in this work. For comparison, we show in dashed lines the redshift distributions from the SOM calibration with lensing weight, which are used in the standard KiDS-1000 cosmological analyses. Coloured bands show the $z_\mathrm{B}$ ranges of corresponding tomographic bins. Dotted lines are $\mathrm{d}I_{\nu}/\mathrm{d}z$ at 353, 545, and 857 GHz calculated from Eq.~\eqref{eq:jsfrd} with the best-fit parameters in this work. The values follow the right $y$ axis.}
    \label{fig:dndz}
\end{figure}

\begin{table}
\caption{Information on the KiDS galaxy sample in each tomographic bin.}
\centering 
\begin{tabular}{llll}\toprule
Bin & $z_{\mathrm{B}}$ range  & Mean redshift  & $\bar{n}[\mathrm{arcmin^{-2}}]$ \\\midrule
1   & {(}0.1, 0.3] & 0.23           & 1.69  \\
2   & {(}0.3, 0.5] & 0.38           & 3.49  \\
3   & {(}0.5, 0.7] & 0.54           & 5.94  \\
4   & {(}0.7, 0.9] & 0.77           & 4.36  \\
5   & {(}0.9, 1.2] & 0.96           & 5.06  \\\bottomrule
\end{tabular}

\label{table:tomoinfo}
\end{table}

We perform a tomographic cross-correlation measurement by dividing the galaxy catalogue into five bins, split according to the best-fit photometric redshift estimate  $z_{\mathrm{B}}$ of each galaxy. These are the same tomographic bins used in the KiDS-1000 cosmology analyses \citep{asgari2020kids1000, heymans2020kids1000, troester2020kids1000}. The redshift distribution of each bin is calibrated using a self-organising map (SOM) as described in \citet{2020A&A...637A.100W, hildebr2020kids1000}. A SOM is a 2D representation of an $n$-dimensional manifold, computed using unsupervised machine learning. For redshift calibration, the SOM classifies the distribution of galaxies in multi-dimensional colour-magnitude space into discrete cells. As galaxy colour correlates with redshift, cells on the SOM similarly correlate with redshift. Using tracer samples of galaxies with known spectroscopic redshift, the associations between galaxies and cells in a SOM can therefore be used to reconstruct redshift distributions for large samples of photometrically defined galaxies. We note that the SOM-calibrated redshift distributions in this study are not the same as \citet{hildebr2020kids1000}, in which the redshift distributions are calibrated with a galaxy sample weighted by the \emph{Lens}fit weight. In this work the redshift distributions are calibrated with the raw, unweighted sample. The redshift distributions of the five tomographic bins are shown in Fig. \ref{fig:dndz}. We also plot the SOM-calibrated $\Phi_{\mathrm{g}}(z)$ with lensing weight as dashed lines. The {absolute} difference between the means of the two $\Phi_{\mathrm{g}}(z)$ are at a level of $\sim 0.01$, comparable to the mean $\Phi_{\mathrm{g}}(z)$ bias given by \citet{hildebr2020kids1000}, and the difference is more evident in the higher-redshift bins. We also show the mean CIB emissions (dotted lines) at 353, 545, and 857 GHz calculated from Eq.~\eqref{eq:jsfrd} with the best-fit parameters in this work.  

We utilised maps of relative galaxy overdensity to encode the projected galaxy density fluctuations. These galaxy overdensity maps are produced for each tomographic bin in the \healpix \citep{Gorski_2005} format with $\textsc{nside} = 1024$, corresponding to a pixel size of 3.4 arcmin. For the $t$-th tomographic bin, the galaxy overdensity in the $i$-th pixel is calculated as
\begin{equation}
    \Delta_{\mathrm{g}, t, i} = \frac{n_{t,i}-\bar{n}_{\mathrm{KiDS},t}}{\bar{n}_{\mathrm{KiDS},t}}
    \label{eq:delta_g_map}
,\end{equation}
where $i$ denotes the pixel index, $n_{t,i}$ is the surface number density of galaxies in the $i$-th pixel and $\bar{n}_{\mathrm{KiDS},t}$ is the mean galaxy surface number density in the $t$-th redshift bin of the KiDS footprint. {We note that Eq.~\eqref{eq:delta_g_map} is slightly different from Eq.~\eqref{eq:delta_g} in that it is the mean galaxy overdensity in each pixel while Eq.~\eqref{eq:delta_g} defines the galaxy overdensity on each point in the sky. In other words, $\Delta_{\mathrm{g}, i}$ in Eq.~\eqref{eq:delta_g_map} is the discretised $\Delta_{\mathrm{g}}(\hat{\theta})$ with the window function corresponding to the \healpix pixel.}
The mask of the galaxy maps for the cross-correlation measurement is the KiDS footprint, which is presented as the transparent regions in Fig. \ref{fig:masks}.

\subsection{CIB data}

In this work, we use the large-scale CIB maps generated by \citet{lenz_large-scale_2019} from three \planck\, High Frequency Instrument (HFI) sky maps at 353, 545, and 857 GHz (the L19 CIB map hereafter)\footnote{\href{https://github.com/DanielLenz/PlanckCIB}{https://github.com/DanielLenz/PlanckCIB}}. The IR signal generated from galactic dust emission is removed based on an HI column density template \citep[][]{2011planckcib, 2014planckxxx}. We use the CIB maps corresponding to an HI column density threshold of $2.0\times 10^{20}\mathrm{cm}^{-2}$. {The CIB mask is a combination of the \planck\, 20\% Galactic mask, the \planck\, extragalactic point source mask, the molecular gas mask, and the mask defined by the HI threshold.} The CIB maps have an overall sky coverage fraction of 25\%, and overlap with most of the KiDS south field and part of the north field. The overlapping field covers about 1\% of the sky. The CIB signal in the maps is in the unit of $\mathrm{MJy/sr}$ with an angular resolution of 5 arcmin, as determined by the full width at half maximum (FWHM) of the beam. The original maps are in the \textsc{Healpix} format with \textsc{Nside}=2048 and we degrade them into \textsc{Nside}=1024 since we do not probe scales smaller than $\ell\sim 1500$.

The \planck\, collaboration also makes all-sky CIB maps \citep{2016planckcib} in the three highest HFI frequencies. To make the all-sky CIB map, the \planck\, collaboration disentangle the CIB signal from the galactic dust emission with the GNILC method \citep{Remazeilles_2011}. These maps have a large sky coverage (about 60\%) and have been extensively used to constrain the CIB power spectra \citep[][]{mak2017measurement,reischke2020information} and to estimate systematics for other tracers \citep{Yan_2019, chluba2017rethinking}. However, \citet{maniyar_star_2018, lenz_large-scale_2019} point out that when disentangling galactic dust from CIB, there is some leakage of the CIB signal into the galactic dust map, causing biases of up to $\sim 20\%$ in the CIB map construction. Therefore, we opt to {not} use the \planck\, GNILC CIB map in this work {at the expenses of sky coverage}.

\subsection{External SFRD data}

In addition to the CIB-galaxy cross-correlation power spectra we also introduce external SFRD measurements, estimated over a range of redshifts, as additional constraints to our model. The external SFRD measurements are obtained by converting the measured IR luminosity functions to $\rho_{\mathrm{SFR}}$ with proper assumptions of the IMF. We refer the interested reader to a review on converting light to stellar mass \citep{madau_cosmic_2014}. In this work, we use the $\rho_{\mathrm{SFR}}$ from \citep[][]{2013ruppioni,2013magnelli, 2016MNRAS.456.1999M, 2016davies}. {We follow \citet{maniyar_star_2018} to account for 
different background cosmologies utilised by these studies: we first convert the external $\rho_{\mathrm{SFR}}$ values into cosmology-independent observed intensity between 8-1000 $\mu \mathrm{m}$ per redshift bin, according to corresponding cosmologies, and then convert back to $\rho_{\mathrm{SFR}}$ with the cosmology assumed in this study.}

\section{Measurements and analysis}
\label{sect:measurements}
\subsection{Cross-correlation measurements}
\label{subsect:ccmeasurements}
The cross-correlation between two sky maps, which are smoothed with the beam window function $b_{\mathrm{beam}}(\ell)$, is related to the real $C_{\ell}$ as
\begin{equation}
    \hat{C}^{uv}_{\ell} = C^{uv}_{\ell}b^u_{\mathrm{beam}}(\ell)b^v_{\mathrm{beam}}(\ell)b^u_{\mathrm{pix}}(\ell)b^v_{\mathrm{pix}}(\ell)
    \label{eq:pseudo_cl}
,\end{equation}
where $\hat{C}^{uv}_{\ell}$ denotes the smoothed $C_{\ell}$ between the sky maps $u$ and $v$, and $b_{\mathrm{pix}}(\ell)$ is the pixelisation window function provided by the \healpix package. In our analysis, we take the Gaussian {beam} window function that is given by
\begin{equation}
    b_{\mathrm{beam}}(\ell) = \exp \left(-\ell(\ell+1)\sigma^2/2\right) 
,\end{equation}
where $\sigma = \mathrm{FWHM}/{\sqrt{8\ln2}}$. For the KiDS galaxy overdensity map, the angular resolution is much better than the pixel size, so we assume a $\mathrm{FWHM} = 0$ and $b^{\mathrm{g}}_{\mathrm{beam}}(\ell)=1$.

Both the galaxy and the CIB maps are partial-sky. The sky masks mix up modes corresponding to different $\ell$. The mode-mixing is characterised by the mode-mixing matrix, which depends only on the sky mask. We use the \textsc{NaMaster}\citep[][]{2019namaster} \footnote{\href{https://github.com/LSSTDESC/NaMaster}{https://github.com/LSSTDESC/NaMaster}} package to correct mode-mixing and estimate the angular cross-power spectra. \textsc{NaMaster} first naively calculates the cross-correlation between two masked maps. This estimation gives the biased power spectrum, which is called the `pseudo-$C_{\ell}$'. Then it calculates the mode-mixing matrices with the input masks and uses this to correct the pseudo-$C_{\ell}$. The beam effect is also corrected in this process. The measured angular power spectra are binned into ten logarithmic bins from $\ell=100$ to $\ell=2000$. The high limit of $\ell$ corresponds to the \planck\, beam, which has FWHM of 5 arcmin. The low limit is set considering the small sky-coverage of KiDS.

The measurements give a data vector of cross-correlation between five tomographic bins of KiDS and 3 CIB bands, resulting in 15 cross-correlations $C^{\nu\mathrm{g}}_{\ell}:  \mathrm{g}\in\{$bin1, bin2, bin3, bin4, bin5$\}; \nu\in \{$353 GHz, 545 GHz, 857 GHz$\}$. Given the covariance matrix to be introduced in Sect.~\ref{subsect:covmat}, we calculate the square-root of the $\chi^2$ values of the null line ($C^{\nu\mathrm{g}}_{\ell}=0$) and reject the null hypothesis at a significance of $43\sigma$. With these measurements, we are trying to constrain the following model parameters with uniform priors: (i) SFR parameters: $\{\eta_{\mathrm{max}}$, $\log_{10}M_{\mathrm{peak}}$, $ \sigma_{M,0}$, $ \tau\}$. The prior boundaries are given in Table.~\ref{tab:sfr_fit}.
(ii) HOD parameters: $\{\log_{10}M_{0,0}$, $\log_{10}M_{0,p}$, $\log_{10}M_{1,0}$, $\log_{10}M_{1,p}$, $ \log_{10}M_{\mathrm{min},0}$, $\log_{10}M_{\mathrm{min},p}\}$. The prior boundaries are given in Table.~\ref{tab:hod_fit}.
(iii) Amplitudes of the shot noise power spectra:  $\left\{\log_{10}{S^{\nu\mathrm{g}}}\right\}$. Here $S^{\nu\mathrm{g}}$ is in the unit MJy/sr. The prior boundaries are $[-12, 8]$ for all the 15 shot noise amplitudes.

In total, there are 25 free parameters to constrain (see Table \ref{table:params} for a summary). The number of data points is 3 (frequencies)$\times$ 5 (tomographic bins) $\times$ 10 ($\ell$ bins)=150, so the degree-of-freedom is 125.

\begin{table}[ht]
\caption{{Summary of the free parameters, their prior ranges, and the equations that define them. The three blocks of the table correspond to three types of parameters: SFR parameters, HOD parameters, and amplitudes of the shot noise power spectra. Note that the last block actually contains 15 shot noise amplitudes.}}
    \centering
    
\begin{tabular} { l l l}

\toprule
 Parameter & Prior & Definition\\
\hline
{$\eta_{\mathrm{max}}$} & [0, 1] & Eq. \eqref{eq:eta}\\

{$\log_{10} M_{\mathrm{peak}}$} & [11.5, 14] & Eq. \eqref{eq:eta} \\

{$\sigma_{M,0}   $} & (0, 4] & Eq. \eqref{eq:eta}, Eq. \eqref{eq:sigma_m}\\

{$\tau           $} & [0, 3] & Eq. \eqref{eq:eta}, Eq. \eqref{eq:sigma_m}\\
\hline 
{$\lgt M_{\mathrm{min,0}}        $} & [9, 14] & Eq. \eqref{eq:ngal_hod}, Eq. \eqref{eq:mp}\\

{$\lgt M_{0,0}          $} & [9, 14] & Eq. \eqref{eq:ngal_hod}, Eq. \eqref{eq:mp}\\

{$\lgt M_{1,0}          $} & [9, 14] & Eq. \eqref{eq:ngal_hod}, Eq. \eqref{eq:mp}\\

{$\lgt M_{\mathrm{min,p}}       $} & [-5, 5] & Eq. \eqref{eq:ngal_hod}, Eq. \eqref{eq:mp}\\

{$\lgt M_{0,p}          $} & [-5, 5] & Eq. \eqref{eq:ngal_hod}, Eq. \eqref{eq:mp}\\

{$\lgt M_{1,p}          $} & [-5, 5] & Eq. \eqref{eq:ngal_hod}, Eq. \eqref{eq:mp}\\
\hline
{$\left\{\log_{10}{S^{\nu\mathrm{g}}}\right\}          $} & [-12, 8] & Eq. \eqref{eq:cltot}\\

\bottomrule

\end{tabular}

    \label{table:params}
\end{table}

\subsection{Covariance matrix}
\label{subsect:covmat}
\begin{figure}
    \centering
    \includegraphics[width=\columnwidth]{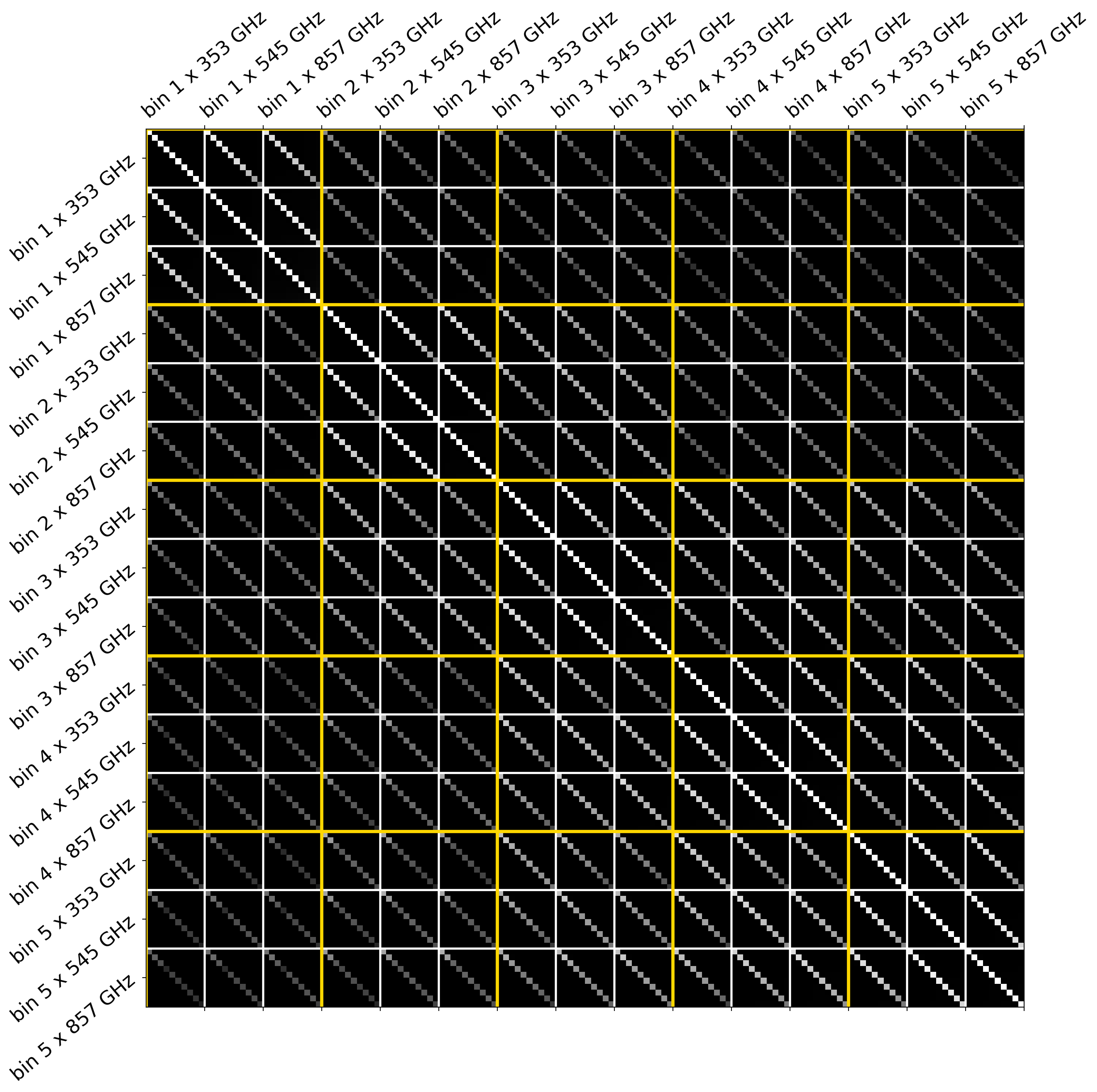}
    \caption{Correlation coefficient matrix of our cross-correlation measurements. The colour scale is from 0 (black) to 1 (white). Each block enclosed by a white grid is the covariance between each pair of cross-correlations indicated with ticks (bin $p \times \nu$ GHz), while that enclosed by a golden grid corresponds to the covariance between the CIB cross-galaxies from each pair of tomographic bins. The matrix has non-zero elements at all cells, but the off-diagonal elements in each cross-correlation are vanishingly small.}
    \label{fig:fullcov}
\end{figure}

To estimate the uncertainty of the cross-correlation measurement, we followed the general construction of the analytical covariance matrix in the literature. Compared with simulation or resampling-based methods, an analytical method is free of sampling noise and allows us to separate different contributions.

Following \citep[][]{troster2021joint}, we decompose the cross-correlation covariance matrix into three parts:

\begin{equation}
    \mathrm{Cov} = \mathrm{Cov^G} + \mathrm{Cov^{T}} + \mathrm{Cov^{SSC}}.
\end{equation}
Here $\mathrm{Cov}$ is the abbreviation of $\mathrm{Cov}_{\ell_1\ell_2}^{uv, wz}\equiv\mathrm{Cov}\left[ C^{uv}_{\ell_1}, C^{wz}_{\ell_2}\right]$. We note that both $\ell_1$ and $\ell_2$ are corresponding to $\ell$ bands rather than a specific $\ell$ mode. The first term $\mathrm{Cov^G}$ is the dominant `disconnected' covariance matrix corresponding to Gaussian fields, including physical Gaussian fluctuations and Gaussian noise:

\begin{equation}\operatorname{Cov}^{\mathrm{G}}\left(C_{\ell_1}^{u v}, C_{\ell_2}^{w z}\right)=\delta_{\ell_1 \ell_2} \frac{C_{\ell_1}^{u w} C_{\ell_2}^{v z}+C_{\ell_1}^{u z} C_{\ell_2}^{v w}}{(2 \ell_1+1)}
\label{eq:covg}
.\end{equation}
This is the covariance matrix for an all-sky measurement. Sky masks introduce non-zero coupling between different $\ell$ as well as enlarge the variance. To account for this, we used the method given by \citet{Efstathiou_2004} and \citet{Garc_a_Garc_a_2019} that is implemented in the \textsc{NaMaster} package \citep{2019namaster}. The angular power spectra in {Eq.~}\eqref{eq:covg} are directly measured from maps so the contribution from noise is also included. We assume that the random noise in the map is Gaussian and independent of the signal.

The second term $\mathrm{Cov^T}$ is the connected term from the trispectrum, which is given by
\begin{equation}
\begin{aligned}
\operatorname{Cov}^{\mathrm{T}}\left(C_{\ell_1}^{u v}, C_{\ell_2}^{w z}\right)&=\int_0^{\infty} \mathrm{d} \chi \frac{W^{u}(\chi) W^{v}(\chi) W^{w}(\chi) W^{z}(\chi)}{4 \pi f_{\mathrm{sky}} \chi^{6}}\\
&\times T_{uvwz}\left(k_1=\frac{\ell_1+1 / 2}{\chi}, k_2=\frac{\ell_2+1 / 2}{\chi}, \chi \right),
\end{aligned}
\end{equation}
where $T_{uvwz}(k)$ is the trispectrum. Using the halo model, the trispectrum is decomposed into one- to four-halo terms. \citet{2018cibng} shows that the one-halo term dominates the CIB trispectrum. As galaxies have a similar spatial distribution to the CIB, we only take the one-halo term into account for our CIB-galaxy cross-correlation:
\begin{equation}
\begin{aligned}
T_{uvwz}^{\mathrm{1h}}(k_1, k_2) &\equiv \int_0^{\infty} \dr M \frac{\dr n}{\dr M} \\
&\times\langle p_u(k_1\mid M) p_v(k_1 \mid M) p_w(k_2\mid M) p_z(k_2\mid M)\rangle.
\end{aligned}
\end{equation}
We will see that this term is negligible in the covariance matrix.

The third term $\mathrm{Cov^{SSC}}$ is called the super sample covariance \citep[SSC;][]{Takada_2013}, which is the sample variance that arises from modes that are larger than the survey footprint. The SSC can dominate the covariance of power spectrum estimators for modes much smaller than the survey footprint, and includes  contributions from halo sample variance, beat coupling, and their cross-correlation. The SSC can also be calculated in the halo model framework \citep[][]{2018ssc, 2021ssc}.

In this work, the non-Gaussian covariance components  $\mathrm{Cov^{T}}$ and $\mathrm{Cov^{SSC}}$ are calculated with the halo model formalism as implemented in {the \textsc{CCL} package} \citep[][]{Chisari_2019}\footnote{\href{https://github.com/LSSTDESC/CCL}{https://github.com/LSSTDESC/CCL}}, and are then summed with $\mathrm{Cov^G}$ to get the full covariance matrix. Unlike \citet{2021yanz}, who calculated covariance matrices independently for the different tomographic bins, the CIB-galaxy cross-correlation in this work is highly correlated across galaxy tomographic bins and CIB frequency bands. Therefore, we calculate the whole covariance matrix of all the 15 cross-correlations, giving a matrix with a side-length of $5\times 3\times 10=150$.

We note that, to calculate the analytical covariance matrix, we needed to use model parameters that we do not know {a priori}. So we utilised an iterative covariance estimate \citep{Ando_2017}: we first took reasonable values for the model parameters, given by \citetalias{maniyar_simple_2021}, to calculate these covariance matrices, and used them to constrain a set of best-fit model  parameters. We then updated the covariance matrix with these best-fit parameters and fitted the model again. In practice, the constraint from the second step is always consistent with the first round, but we nonetheless took the constraints from the second step as our fiducial results.

Figure \ref{fig:fullcov} shows the correlation coefficient matrix. The five diagonal golden blocks have very high off-diagonal terms, which means that the cross-correlations between galaxies in the same tomographic bin with three CIB channels have a very high correlation (about $95\%$). This is because the CIB signals from different frequencies are essentially generated by the same galaxies. The correlation of off-diagonal golden blocks is weaker but still non-negligible: this correlation is from the overlap of galaxy redshift distributions in different tomographic bins, as shown in Fig. \ref{fig:dndz}. We also note that the SSC term contributes up to $8\%$ of the total standard deviation, while the trispectrum term is insignificant (contributing $<0.1\%$ to all covariance terms). This is in contrast to \citet{troster2021joint}, who study tSZ cross-correlations and find that the SSC term was a more significant contributor to their covariance (contributing $\sim 20\%$ to their off-diagonal covariance terms). The reason for this difference is that, compared to the tSZ effect, the galaxy distribution is more concentrated. This causes the non-Gaussian term to remain insignificant until considerably smaller scales than the tSZ effect: beyond the scales probed in this study ($\ell>2000$).

{Finally, an alternative estimation of the covariance matrix is shown in Appendix~\ref{app:jackcov}}.

\subsection{Systematics}

\subsubsection{CIB colour-correction and calibration error}

The flux given in the \planck\ sky maps follows the photometric convention that $\nu I_{\nu}$=constant \citep{2014planckxxx}. The flux therefore has to be colour-corrected for sources with a different SED. Therefore, the CIB-galaxy cross-correlation should also be corrected as
\begin{equation}
    C^{\nu\mathrm{g},\mathrm{measured}}_{\ell} = C^{\nu\mathrm{g},\mathrm{model}}_{\ell}\times cc_{\nu},
\end{equation}
where $cc_{\nu}$ is the colour-correction factor at frequency $\nu$. In this work, we adopt the colour-correction factors given by \citet{bethermin2012unified} with values of 1.097, 1.068, and 0.995 for the 353, 545, and 857 GHz bands, respectively.

Additionally, in \citet{maniyar_star_2018, maniyar_simple_2021} the authors introduce a scaling factor as an additional calibration tool when working with the L19 CIB maps. However, they constrain this factor to be very close to one (at a level of $\sim \pm 1\%$). As such, in this work we neglect the additional calibration factor.

\subsubsection{Cosmic magnification}
The measured galaxy overdensity depends not only on the real galaxy distribution but also on lensing magnification induced by the line-of-sight mass distribution \citep{1989A&A...221..221S, 1989ApJ...339L..53N}. This so-called cosmic magnification has two effects on the measured galaxy overdensity: i) overdensities along  the line-of-sight cause the local angular separation between source galaxies to increase, so the galaxy spatial distributions are diluted and the cross-correlation is suppressed; and ii) lenses along the line-of-sight magnify the flux of source galaxies such that fainter galaxies enter the observed sample, so the overdensity typically increases. These effects potentially bias galaxy-related cross-correlations, especially for high-redshift galaxies \citep{hui2007anisotropic, ziour2008magnification, hilbert2009ray}. Cosmic magnification depends on the magnitude limit of the galaxy survey in question, and on the slope of the number counts of the sample under consideration at the magnitude limit. We follow \citet{2021yanz} to correct this effect, and note that this correction has a negligible impact on our best-fit CIB parameters.

\subsubsection{Redshift distribution uncertainty}

The SOM-calibrated galaxy redshift distributions have an uncertainty on their mean at a level of $\sim 0.01$ \citet{hildebr2020kids1000}, which could affect galaxy cross-correlations. To test the robustness of our results to this uncertainty, we run a further measurement including additional free parameters that allow for shifts in the mean of the redshift distributions. With these additional free parameters, the shifted galaxy redshift distributions are given by
\begin{equation}
    \tilde{\Phi}_{\mathrm{g},i}(z) = {\Phi}_{\mathrm{g},i}(z+\delta_{z,i}),
\end{equation}
where $\tilde{\Phi}_{\mathrm{g},i}(z)$ is the shifted galaxy redshift distribution in the $i^\mathrm{th}$ tomographic bin, and  $\delta_{z,i}$ is the shift of the redshift distribution parameter in the $i^\mathrm{th}$ bin. The priors on $\delta_{z,i}$ are assumed to be covariant Gaussians centred at $\bar{\delta}_{z,i}$ (i.e. the mean $\delta_{z,i}$). \citet{hildebr2020kids1000} evaluated both $\bar{\delta}_{z, i}$ and the covariance matrix from simulations, but did so using the redshift distributions calculated with lensing weights. As previously discussed, however, the $\Phi_{\mathrm{g}}(z)$ used in this work is from the SOM calibration without lensing weights. Therefore, neither the estimates of $\bar{\delta}_{z,i}$ nor their covariance matrix from \citet{hildebr2020kids1000} are formally correct for the $\Phi_{\mathrm{g}}(z)$ in this work. To correctly estimate  $\bar{\delta}_{z,i}$ and the associated covariance matrix for this work, one needs to analyse another simulation suite for the SOM calibration without lensing weight, which is not currently  available. However, given the similarity between the lensing-weighted and unweighted redshift distributions (Fig.~\ref{fig:dndz}), we can alternatively adopt a conservative prior on $\delta_{z,i}$ with the mean values and uncertainties that are three times than the fiducial values given by \citet{hildebr2020kids1000}. Our fiducial $\delta_{z,i}$ covariance matrix is therefore simply defined as nine times the nominal KiDS-1000 $\delta_{z,i}$ covariance matrix. This yields an {absolute} uncertainty at a level of 0.04, about two times the difference between the nominal KiDS-1000 lensing-weighted $\Phi_{\mathrm{g}}(z)$ and the unweighted $\Phi_{\mathrm{g}}(z)$ that we use in this work. 

\subsection{Likelihood}

We constrained the model parameters in two ways. The first is cross-correlation only (`\cco fitting' hereafter), and the second is combining cross-correlation and external $\rho_{\mathrm{SFR}}$ measurements (`\ccosfrd fitting' hereafter). For the cross-correlation only fitting, since we are working with a wide $\ell$ range, there are many degrees-of-freedom in each $\ell$ bin. According to the central limit theorem, the bin-averaged $C_{\ell}$'s obey a Gaussian distribution around their true values. Thus, we assume that the measured power spectra follow a Gaussian likelihood:

\begin{equation}
    -2 \ln L(\boldsymbol{\tilde{C}} \mid\vec{q})=\chi^{2} \equiv(\boldsymbol{\tilde{C}}-\boldsymbol{C}(\vec{q}))^{T} \mathrm{Cov}^{-1}(\boldsymbol{\tilde{C}}-\boldsymbol{C}(\vec{q}))
,
\label{eq:like}
\end{equation}
where $\vec{q}$ stands for our model parameters given in Sect.~\ref{sect:measurements}. For the additional chain to test the robustness to redshift uncertainties, $\vec{q}$ also includes $\delta_{z,i}$ introduced in the previous subsection. The data vector $\boldsymbol{\tilde{C}}$ is a concatenation of our 15 measured CIB-galaxy cross-correlations; $\boldsymbol{C}(\vec{q})$ is the cross-correlation predicted by the model described in Sect.~\ref{sect:model} with parameters $\vec{q}$. 

The external $\rho_{\mathrm{SFR}}$ measurements are assumed to be independent of our cross-correlation and similarly independent at different redshifts. Therefore, including them introduces an additional term in the likelihood:
\begin{equation}
\begin{aligned}
      -2 \ln L(\{\boldsymbol{\tilde{C}}, \tilde{\rho}_{\mathrm{SFR}}\} \mid\vec{q})&=-2 \ln L(\boldsymbol{\tilde{C}} \mid\vec{q}) \\
     & + \sum_i\frac{ \left[\tilde{\rho}_{\mathrm{SFR}}(z_i)- {\rho}_{\mathrm{SFR},i}(z_i|\vec{q})\right]^2}{\sigma_{\rho, i}^2}
,  
\end{aligned}
\label{eq:like_esfr}
\end{equation}
where $\tilde{\rho}_{\mathrm{SFR}}(z_i)$ is the measured SFRD at the redshift $z_i$, while ${\rho}_{\mathrm{SFR},i}(z_i|\vec{q})$ is that predicted by our model (see Eq.~\eqref{eq:sfrd_z}), and $\sigma_{\rho, i}$ is the standard error of these SFRD measurements. {We note that, we are still constraining the same HOD measurement as the cross-correlation-only constraint.}

We also perform fitting with only external ${\rho}_{\mathrm{SFR}}$ data and errors. This test serves as a consistency check between the \cco measurement and previous multi-wavelength studies, as well as a validation of our CIB halo model. The free parameters are the SFR parameters and HOD parameters (ten parameters in total).

We adopt the Markov-Chain Monte-Carlo method to constrain our model parameters with the python package \textsc{emcee} \citep[][]{Foreman_Mackey_2013}. Best fit parameters are determined from the resulting chains, as being the sample with the smallest $\chi^2$ goodness-of-fit. Marginal constraints on parameters, when quoted, are marginal means and standard deviations.

\section{Results}
\label{sect:results}

\subsection{Constraints on star formation rate density}
\label{subsect:sfrd}

\begin{figure*}
    \centering
    \includegraphics[width=\textwidth]{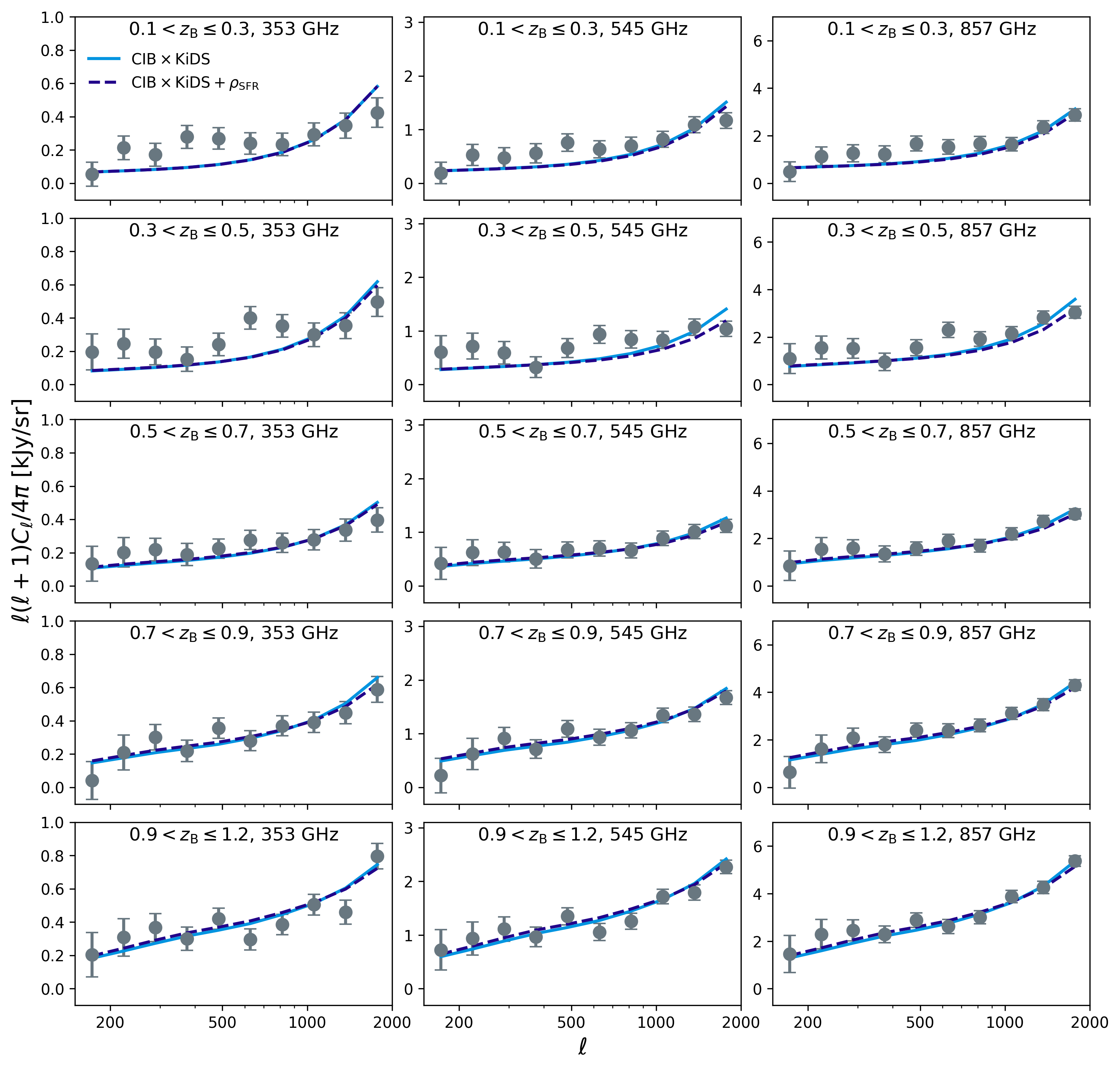}
    \caption{CIB-galaxy cross-correlations with the five KiDS tomographic bins (rows) and the three CIB maps (columns). The grey points are measured from data, with standard deviation error bars calculated using the square root of the diagonal terms of the covariance matrix. The solid cyan lines show the best-fit cross-correlation signals calculated using the CIB-galaxy cross-correlation measurements alone, while the dashed blue lines show the best-fit cross-correlations when jointly fitting the CIB-galaxy cross-correlation measurements and the external SFRD. }
    \label{fig:kc_cc}
\end{figure*}

\begin{figure*}
    \centering
    \includegraphics[width=0.8\textwidth]{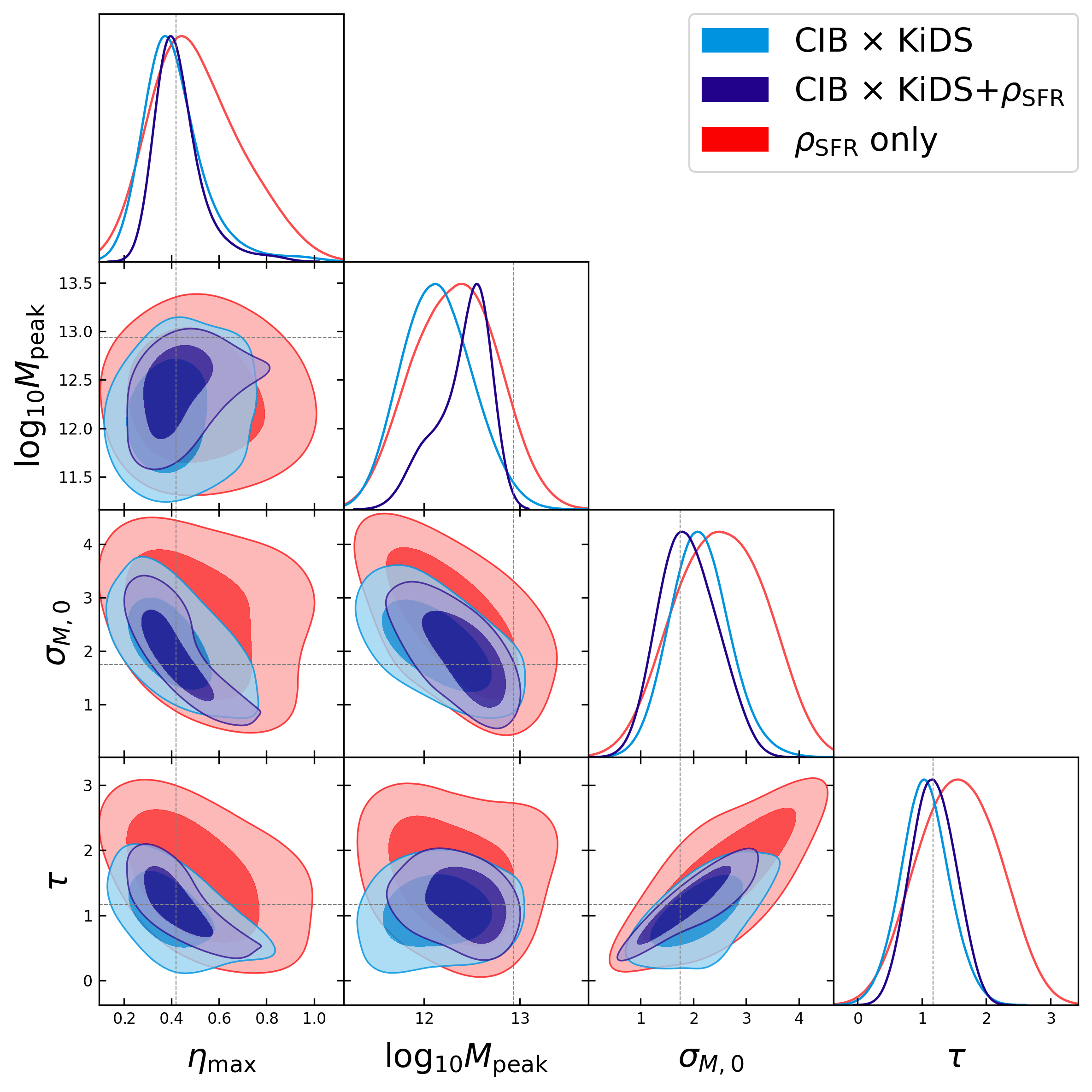}
    \caption{{Posterior of the SFR parameters. Contours show the 2D posteriors marginalised over all the other 25-2=23 parameters.} The cyan contours show the constraints from the CIB-KiDS cross-correlation only, the dark blue contours show the constraints from a combination of the cross-correlation and the external SFRD data, and the red contours show the constraints from the external SFRD data only. {The dark and light regions in each contour show the 68\% and 95\% credible regions, respectively.} The dashed lines show the best-fit model from \citetalias{maniyar_simple_2021}.}
    \label{fig:mcmc_post}
\end{figure*}

\begin{table*}[ht]
\caption{Summary of the prior ranges, the marginalised mean values, and the 68\% {credible intervals} of SFR parameters. {The values and errors are calculated from the posteriors marginalised over all the other parameters.} The last column shows the constraint from \citetalias{maniyar_simple_2021} as a reference. The last two rows show the $\chi^2$ values, degrees-of-freedom, and PTE for our constraints.}
    \centering
    
\begin{tabular} { l l c c c c}
\toprule
 Parameter & Prior & \cco & \ccosfrd & $\rho_{\mathrm{SFR}}$ only & \citetalias{maniyar_simple_2021}\\
\hline
{$\eta_{\mathrm{max}}$} & [0, 1] & $0.41^{+0.09}_{-0.14}$  & $0.427^{+0.065}_{-0.11}    $ & $0.51^{+0.16}_{-0.22}$ & $0.42^{+0.03}_{-0.02}$\\

{$\log_{10} M_{\mathrm{peak}}$} & [11.5, 14] & $12.14^{+0.36}_{-0.36}$  & $12.42^{+0.35}_{-0.19}$ & $12.33^{+0.42}_{-0.42}$ & $12.94^{+0.02}_{-0.02}$ \\

{$\sigma_{M,0}   $} & (0, 4] & $2.11^{+ 0.55}_{-0.55}$ & $1.91^{+0.51}_{-0.61}$ & $2.52^{+0.82}_{-0.82}$ & $1.75^{+0.12}_{-0.13}$\\

{$\tau           $} & [0, 3] & $1.05^{+0.37}_{-0.37}$ & $1.18^{+0.34}_{-0.34}$ & $1.57^{+0.61}_{-0.61} $ & $1.17^{+0.09}_{-0.09}$\\
\hline 
$\chi^2$/d.o.f & - & 142.82/125=1.14 & 155.99/142=1.10 &5.76/7=0.82 & -\\
PTE & - & 0.13 & 0.21 & 0.57 & -\\
\bottomrule
\end{tabular}

    \label{tab:sfr_fit}
\end{table*}

We show our CIB-KiDS cross-correlation measurement in Fig. \ref{fig:kc_cc}. Each panel presents the cross-correlation between galaxies from a tomographic bin and CIB anisotropies in one frequency band. The points are the mean $C_{\ell}$ in each of the ten logarithmic $\ell$ bins. The error bars are the standard deviation calculated as the square root of the diagonal terms of the covariance matrix. We show the cross-correlations calculated from the model given by Sect.~\ref{sect:model} from the constrained parameters with \cco fitting (see Table~\ref{tab:sfr_fit}) as well as those calculated from constrained parameters given by the \ccosfrd fitting. The reduced $\chi^2$ (RCS) for both fits are 1.14 and 1.10, with degrees-of-freedom 125 and 142, respectively. In order to evaluate the goodness-of-fit, we calculate the corresponding probability-to-exceed (PTE) given the degree-of-freedom: our fits give PTE values of $0.13$ and $0.2$ for the \cco and \ccosfrd fitting, respectively. \citet{heymans2020kids1000} adopts the criterion PTE>0.001 (corresponding to a $\sim3\sigma$ deviation) to be acceptable. Therefore, we conclude that our model generally fits the cross-correlations well. We also notice that the fitting in low-redshift bins is worse than high-redshift bins (see the `$0.1<z_{\mathrm{B}}\leq0.3$, 353 GHz' panel in Fig.~\ref{fig:kc_cc}, for example), although correlation in the points makes `chi-by-eye' inadvisable here.

We estimate the posterior with the last 100000 points of each of our chains: \cco, \ccosfrd, and $\rho_{\mathrm{SFR}}$-only. The posterior of the four SFR parameters are shown in the triangle plot in Fig. \ref{fig:mcmc_post}. The distributions are marginalised over HOD parameters and shot noise amplitudes. In particular, we note the {good} constraint that our \cco only results have over SFR parameters (cyan contours). The cross-correlation only results are consistent with the results when analysing external SFRD data only (i.e. the red contours). This validates our  CIB-galaxy cross-correlation as a consistent yet independent probe of the cosmic SFRD, and further demonstrates the validity of the halo model (used in both studies) when applied to vastly different observational data. It is also encouraging that the cross-correlation constraints are generally tighter than those made with the external SFRD data alone, demonstrating that the cross-correlation approach provides a valuable tool for studying the cosmic SFR history. Our joint constraints are tighter still, demonstrating that there is different information in the two probes that can be leveraged for additional constraining power in their combination (the \ccosfrd constraints shown in dark blue). The marginal parameter values and uncertainties are shown in Table \ref{tab:sfr_fit}, and are calculated as the mean and 68\% confidence levels of the Gaussian kernel density estimate (KDE) constructed from the marginal posterior samples. The Gaussian kernel is dynamically adapted to the distribution of the marginal samples using the \citet{Botev_2010} diffusion bandwidth estimator. 

To evaluate the constraining power, we adopt the method from \citet{asgari2020kids1000}: we calculate the values of the marginalised posterior at both extremes of the prior distribution, and compare them with 0.135, the ratio between the peak of a Gaussian distribution and the height of the $2\sigma$ confidence level. If the posterior at the extreme is higher than 0.135, then the parameter boundary is not well constrained. We find that except the lower bound of $M_\mathrm{peak}$ for \cco and the higher bound of sigma for $\rho_{\mathrm{SFR}}$-only, the other parameters are all constrained.

\begin{figure}
    \centering
    \includegraphics[width=\columnwidth]{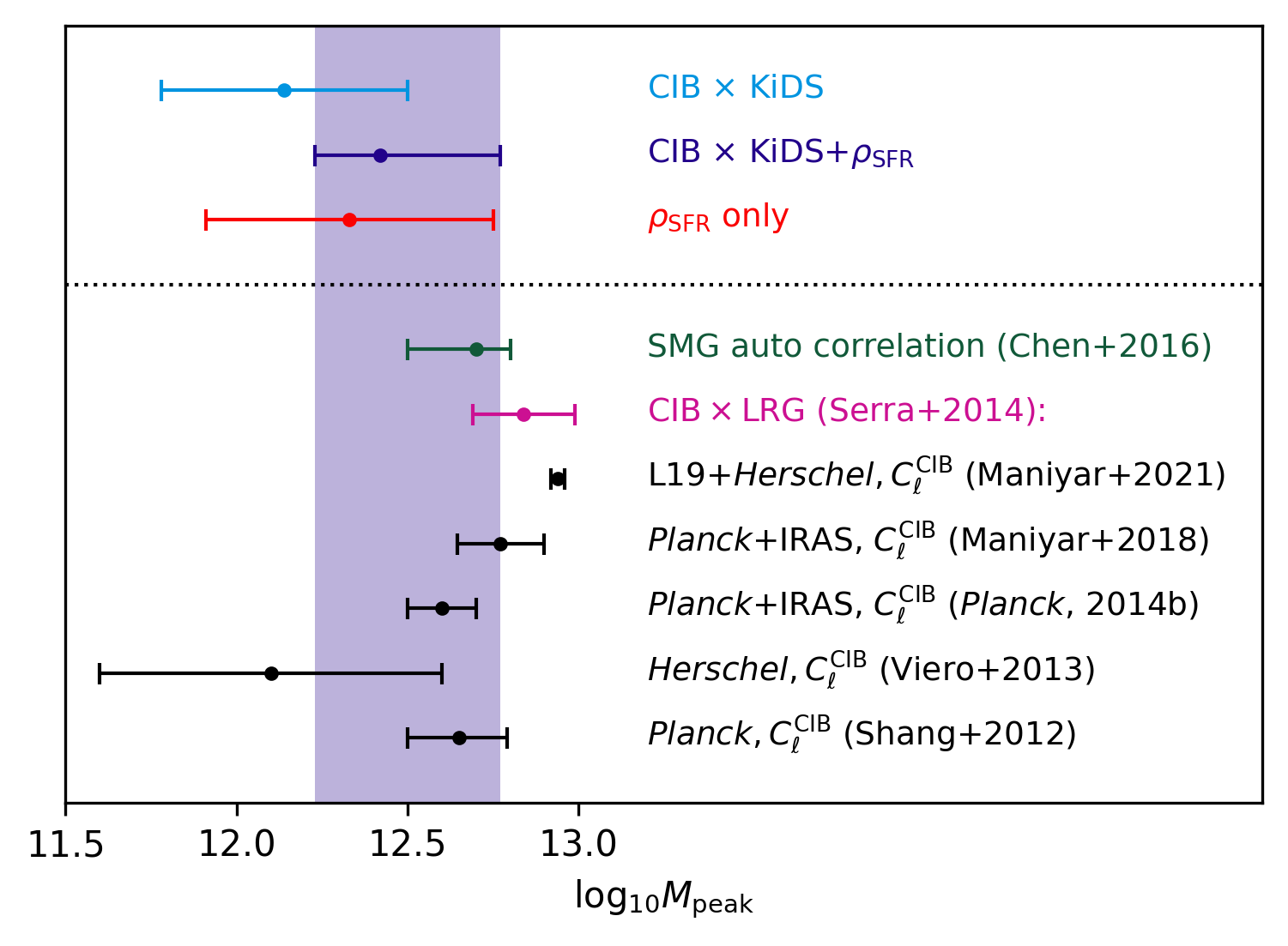}
    \caption{Comparison of our constraints on $M_{\mathrm{peak}}$ with a number of recent results from the literature. The three points above the dotted line are the results from this work. The other points are colour-coded according to their methods: the green point shows the result from SMG auto-correlations \citep[][]{2016ApJ...831...91C}, the magenta shows the measurements using LRG-CIB cross-correlations \citep[][]{Serra_2014}, and the black points show measurements using CIB power spectra  \citep[][]{shang_improved_2012, 2013ApJ...772...77V, 2014planckxxx,  maniyar_star_2018, maniyar_simple_2021}. The dark blue band shows the 68\% credible interval of our \ccosfrd marginal posterior constraint.}
    \label{fig:mpeak_review}
\end{figure}

One of the parameters that is of particular interest is $M_{\mathrm{peak}}$: the halo mass that houses the most efficient star formation activities. In Fig. \ref{fig:mpeak_review}, we summarise our results and recent observational results from the literature that have constrained this parameter. The three points above the dotted line are the constraints from this work. The other points are colour-coded according to their methods: the green point shows the result of SMG auto-correlations \citep[][]{2016ApJ...831...91C}, the magenta point shows the measurement using LRG-CIB cross-correlations \citep[][]{Serra_2014}, and black points show measurements utilising CIB power spectra  \citep[][]{shang_improved_2012, 2013ApJ...772...77V, 2014planckxxx,  maniyar_star_2018, maniyar_simple_2021}. The purple band shows the 68\% credible interval of our \ccosfrd constraint. Except \citetalias{maniyar_simple_2021}, our constraints are in agreement with previous studies within the 2$\sigma$ level, but prefer a slightly lower $M_{\mathrm{peak}}$. This may be due to the different data used in these studies, which would suggest that estimates of  $M_{\mathrm{peak}}$ depend on galaxy types. Our results are in a mild tension with \citetalias{maniyar_simple_2021}, which we hypothesis may be due to an inaccurate uncertainty estimate for their model parameters, driven by their assumption of a purely diagonal covariance matrix.

It is interesting to compare our result with \citetalias{maniyar_simple_2021} because they also constrain SFRD history with the same halo model, but from CIB power spectra. According to Appendix A in \citetalias{maniyar_simple_2021}, without introducing external SFRD data, CIB power spectra yield biased SFRD measurements at low redshift. Without the redshift information from SFRD measurements, {the constraining power of model parameters are limited by degeneracy between them}, which is reasonable because the CIB is a line-of-sight integrated signal. In this regime, all parameters that describe the redshift distribution of CIB emission ($M_{\mathrm{peak}}, \sigma_{M,0}, \tau, z_c$) should be degenerate. Therefore, it is remarkable that our \cco constraints are able {to constrain both $\sigma_{M,0}$ and $\tau$}. We attribute this increased precision to the use of tomography in our \cco measurements.

\begin{figure}
    \centering
    \includegraphics[width=\columnwidth]{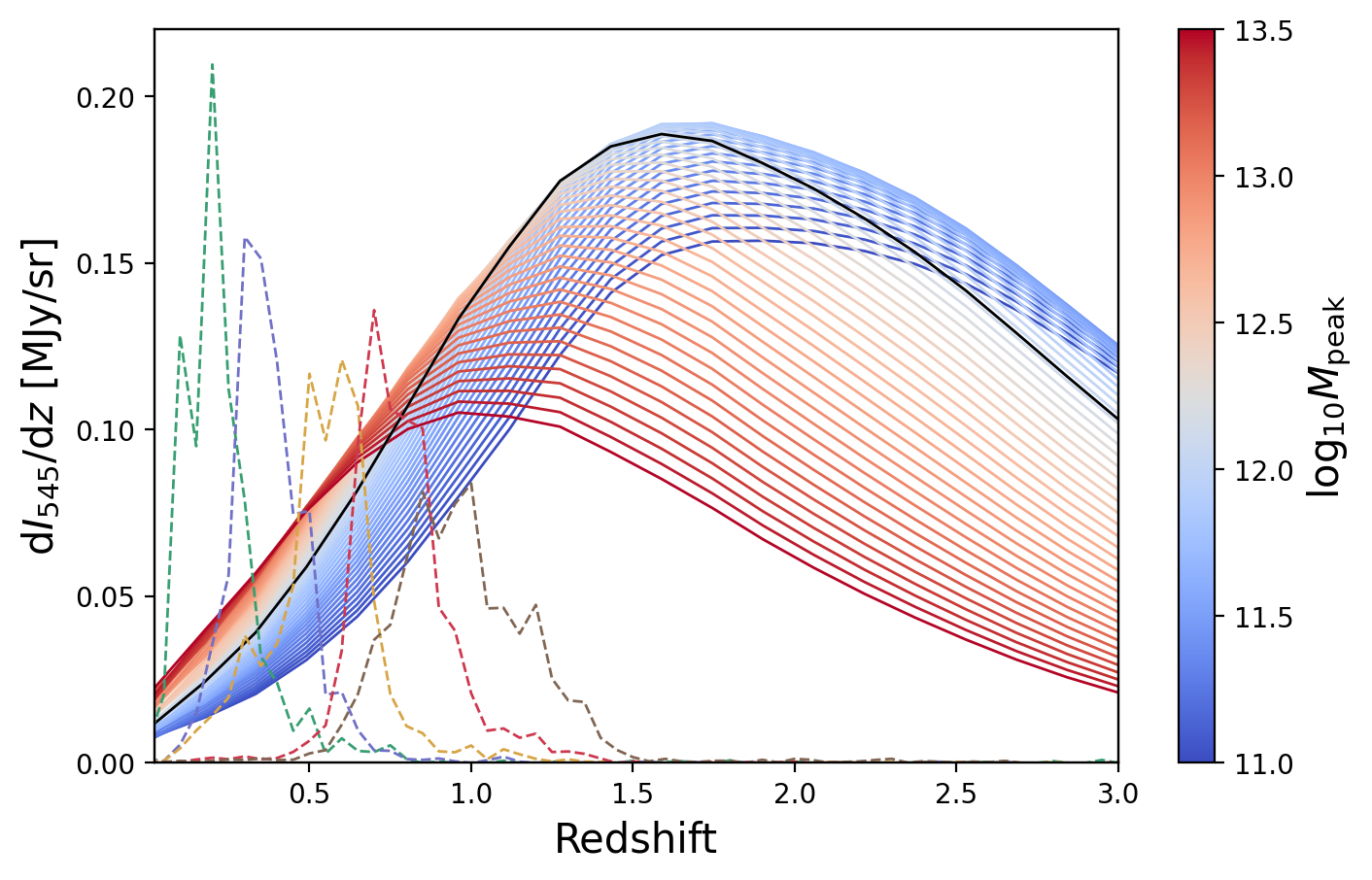}
    \caption{CIB intensity at 545 GHz as a function of redshift while varying $M_{\mathrm{peak}}$ and keeping all other parameters fixed (solid lines, right y axis). We also plot the redshift distributions of the galaxy sample $\Phi_{\mathrm{g}}(z)$ (dashed lines, left y axis). The CIB emissions are calculated from Eq.~\eqref{eq:jsfrd}. The black line shows $\mathrm{d}I_{545}/\mathrm{d}z$, which corresponds to the \cco best-fit parameters.}
    \label{fig:sfrd_varym}
\end{figure}

We note that the cross-correlation-only measurement does not constrain $\log_{10} M_{\mathrm{peak}}$ well. This is because $M_{\mathrm{peak}}$ depends primarily on the CIB signal at high redshifts. We verify this by calculating the redshift distribution of mean CIB intensity defined in Eq.~\eqref{eq:cibintens}, while varying $M_{\mathrm{peak}}$ and fixing all other parameters. The result at 545 GHz is presented in Fig.~\ref{fig:sfrd_varym}; results using the other two channels show similar behaviour. It is clear that varying $M_{\mathrm{peak}}$ affects the CIB signal at high redshift more dramatically than at low redshift. In the redshift range of our galaxy sample, the mean CIB emission does not change significantly with $M_{\mathrm{peak}}$ as much as $z>1.5$, especially for the lowest tomographic bins. Therefore, external SFRD measured at high redshifts, where the CIB intensity is more sensitive to $M_{\mathrm{peak}}$, provides additional constraints on $M_{\mathrm{peak}}$.

\begin{figure*}[t]
\centering

\subfloat[]{%
  \includegraphics[width=\columnwidth]{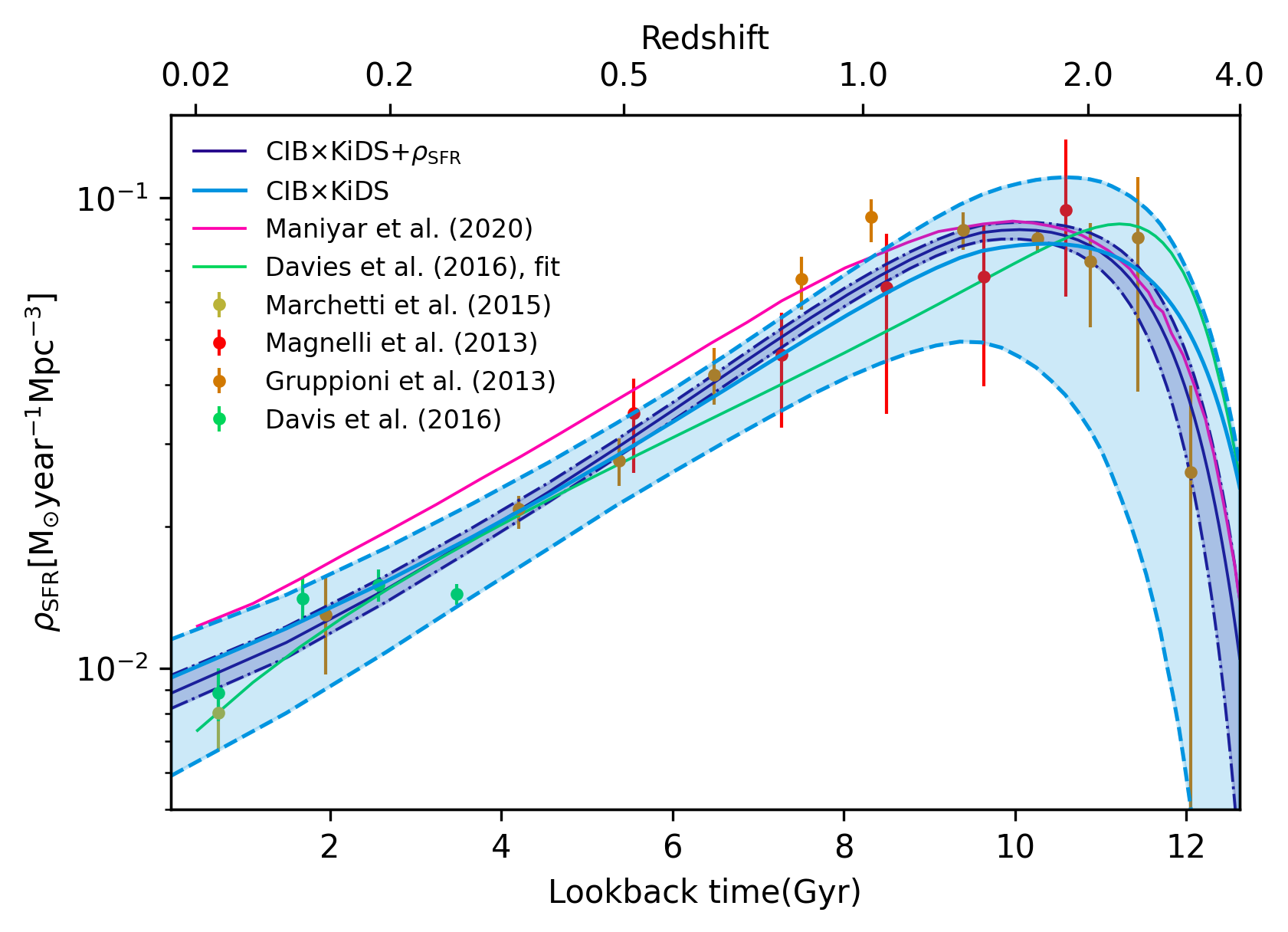}%
  \label{fig:sfrd_t}%
}\qquad\hspace{-1cm}
\subfloat[]{%
  \includegraphics[width=\columnwidth]{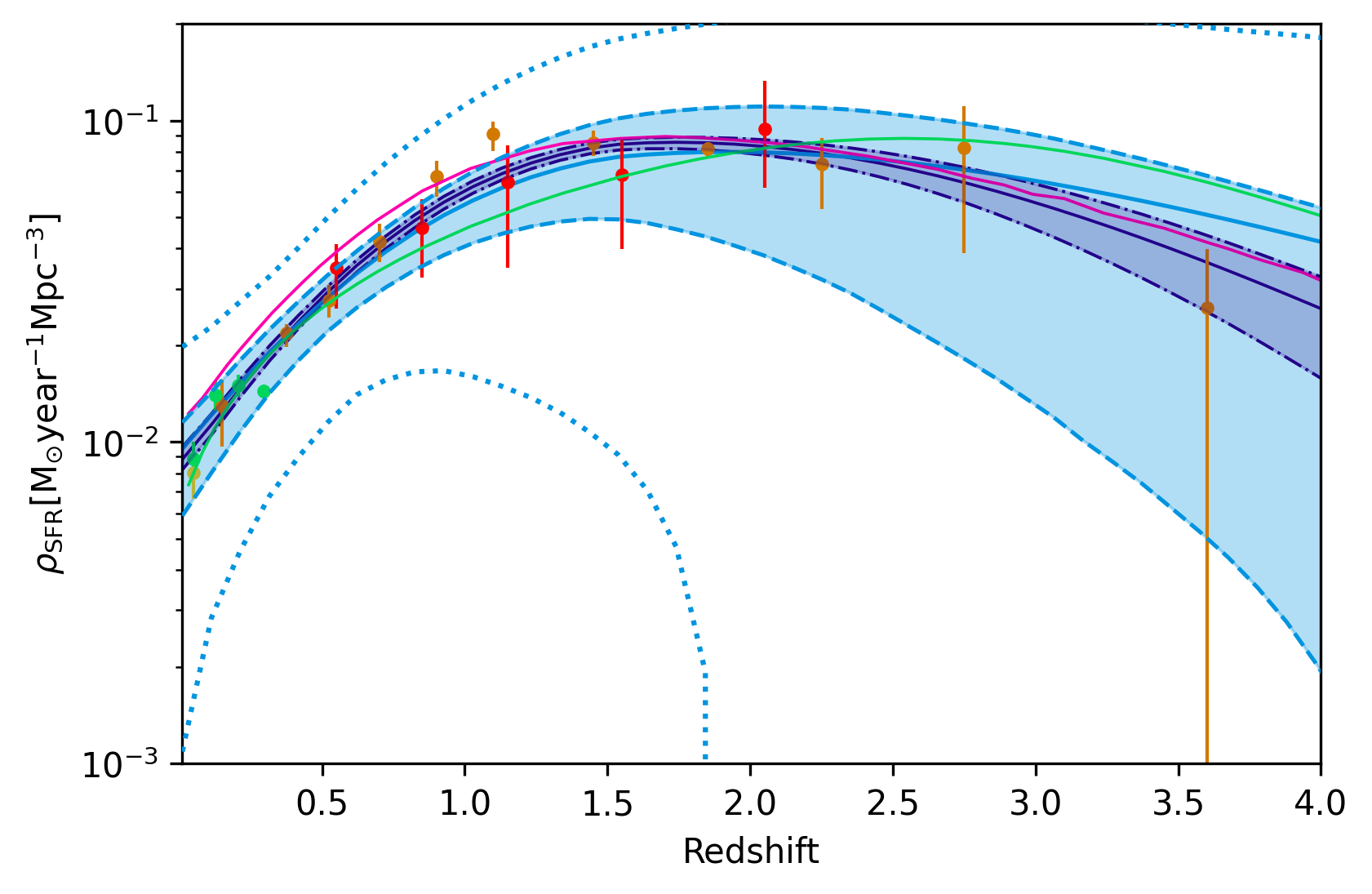}%
  \label{fig:sfrd_z}%
}

\caption{Evolution of the SFRD with respect to lookback time (panel a) and redshift (panel b). The SFRD calculated from this work is presented as cyan (cross-correlation only) and dark blue (cross-correlation plus external SFRD) lines and shaded regions. The shaded regions enclose the $1\sigma$ credible region of the fits and are calculated from 10000 realisations of SFR parameters from the posterior distribution. {The $3\sigma$ credible region of the cross-correlation-only SFRD is also shown in panel b with dotted cyan lines. We note that the lower $3\sigma$ limit crosses zero at $z\sim 1.8$} The magenta and green lines are the best-fit SFRD from two previous studies, and the points with error bars are the SFRD from previous studies (which are included in our $\mathrm{CIB}\times\mathrm{KiDS}+\rho_{\mathrm{SFR}}$ analysis).}
\label{fig:sfrd}
\end{figure*}

The constraints of SFRD is shown in Fig. \ref{fig:sfrd} with respect to lookback time in panel (a) and redshift in panel (b). The \cco fit is shown in cyan line while the \ccosfrd in dark blue. We estimate the credible interval of our fits by calculating SFRD using 10000 realisations of our model parameters, drawn from the posterior distribution shown in Fig \ref{fig:mcmc_post}. The 68\% credible interval is shown as bands with corresponding colours, which are calculated by computing the 10000 samples from the model and deriving the SFRD at a range of redshifts. Credible intervals are computed at each redshift using a Gaussian KDE, and these constraints are connected to form the filled regions. The magenta and green lines are the best-fit SFRD from \citetalias{maniyar_simple_2021} and \citet{2016davies}, and the points with error bars are SFRD estimates from previous studies (which are included in our $\mathrm{CIB}\times\mathrm{KiDS}+\rho_{\mathrm{SFR}}$ analysis). The SFRD fitting given by \citetalias{maniyar_simple_2021} is from a combination of the CIB auto- and cross-correlation power spectra and external SFRD measurements, while \citet{2016davies} is an empirical model estimated using galaxy surveys. These two figures demonstrate that our measurements agree well with previous studies using different analysis techniques. The \ccosfrd fitting agrees well with external SFRD in all redshift ranges. Notably, we are also able to obtain fairly accurate and precise constrain on SFRD up to $z\sim 1.5$ (corresponding to a lookback time of 10 Gyr) using our \cco cross-correlations alone. Beyond $z\sim 1.5$, \cco fitting yields large uncertainties because our KiDS sample has few galaxies beyond this point. {The constraint of SFRD drops below $3\sigma$ level beyond $z\sim1.8$ {(see the dotted lines in Fig.~\ref{fig:sfrd_z})}} As a result, our sample is not deep enough to directly constrain $M_{\mathrm{peak}}$. We conclude that the \cco constraint yields a peak SFRD of $0.08_{-0.04}^{+0.03} \,\mathrm{M}_{\odot} \mathrm { year }^{-1} \mathrm{Mpc}^{-3}$ at $z=1.94^{+0.1}_{-0.51}$, corresponding to a lookback time of $10.42^{+0.16}_{-1.07}$ Gyr, while \ccosfrd constraint gives a peak SFRD of $0.09_{-0.004}^{+0.003}\,\mathrm{M}_{\odot} \mathrm { year }^{-1} \mathrm{Mpc}^{-3}$ at $z=1.74^{+0.06}_{-0.02}$, corresponding to a lookback time of $10.05^{+0.12}_{-0.03}$ Gyr, consistent with previous observations of the cosmic star formation history.

\begin{figure}
    \centering
    \includegraphics[width=\columnwidth]{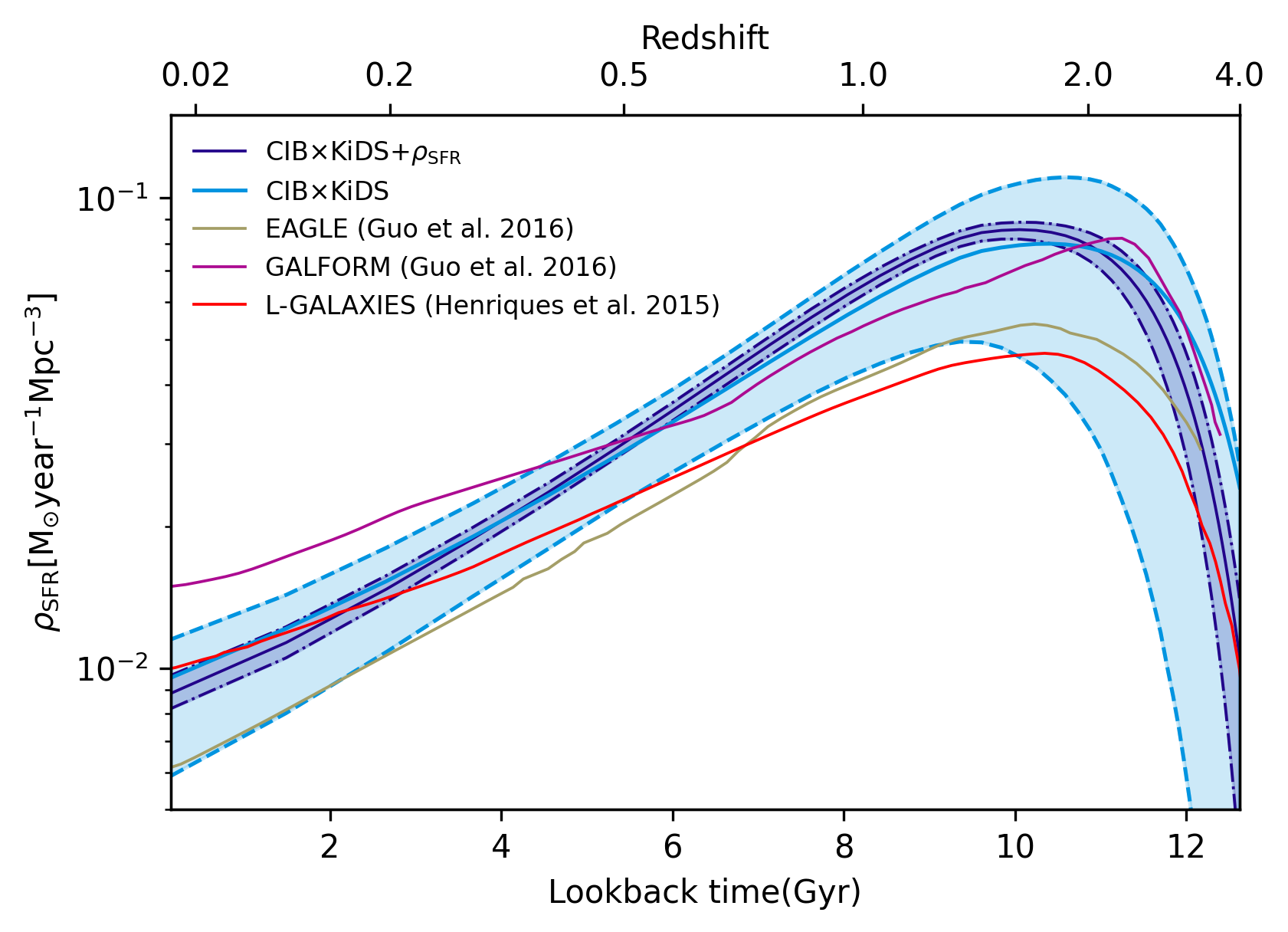}
    \caption{Evolution of the SFRD with respect to lookback time from this work (see Fig. \ref{fig:sfrd}), compared to simulations and SAMs.}
    \label{fig:sfrd_sim}
\end{figure}

{In parallel with observations, simulations and semi-analytical models (SAMs) also give estimations on SFRD. In order to check the consistency between observation, simulations and SAMs, } we compare our constraints on the SFRD with results given by simulations and SAMs in Fig. \ref{fig:sfrd_sim}. {The results are from the \textsc{galform} SAM \citep[][with the \citealt{Gonzalez-Perez+2014_GALFORM} version; purple line]{Guo+2016_Gal}, \textsc{EAGLE} \citep[][with the \citealt{Schaye+2015_EAGLESim} simulation; khaki line]{Guo+2016_Gal}, and the \textsc{L-GALAXIES} SAM \citep[][red line]{Henriques+2015_LGal}. These models adopt different simplifications for active galactic nucleus (AGN) and supernova feedbacks, different star formation thresholds, and different gas-cooling models.} We see that \textsc{EAGLE} predicts a slightly lower SFRD (at essentially all times) than is predicted from our results. \textsc{galform}, on the other-hand, agrees well with our \cco fits at intermediate-to-early times, but predicts a higher SFRD than our model in the 1-5 Gyr range. As is discussed in \citet{2017driver}, this might be due to the high dust abundance at low redshift in the \citet{Gonzalez-Perez+2014_GALFORM} version of \textsc{galform}. \textsc{L-GALAXIES} incorporates different {methodologies to model the environment, cooling processes, and star formation} of galaxies. Combining these modifications ensures that massive galaxies are predominately quenched at earlier times while low mass galaxies have star formation histories that are more extended into the late-time Universe. As can be seen, {this results in a low SFRD prediction} at intermediate redshifts compared to our data, but a better fit at low redshift. However, given the large uncertainties on our \cco fits, the CIB-galaxy cross-correlation is currently not precise enough to invalidate any of these simulations. Comparing to our \ccosfrd analysis, none of the three simulations are able to reproduce our fits at all redshifts. This highlights the complexity of the  underlying physics that determines the form of the observed SFRD.

\begin{figure}
    \centering
    \includegraphics[width=\columnwidth]{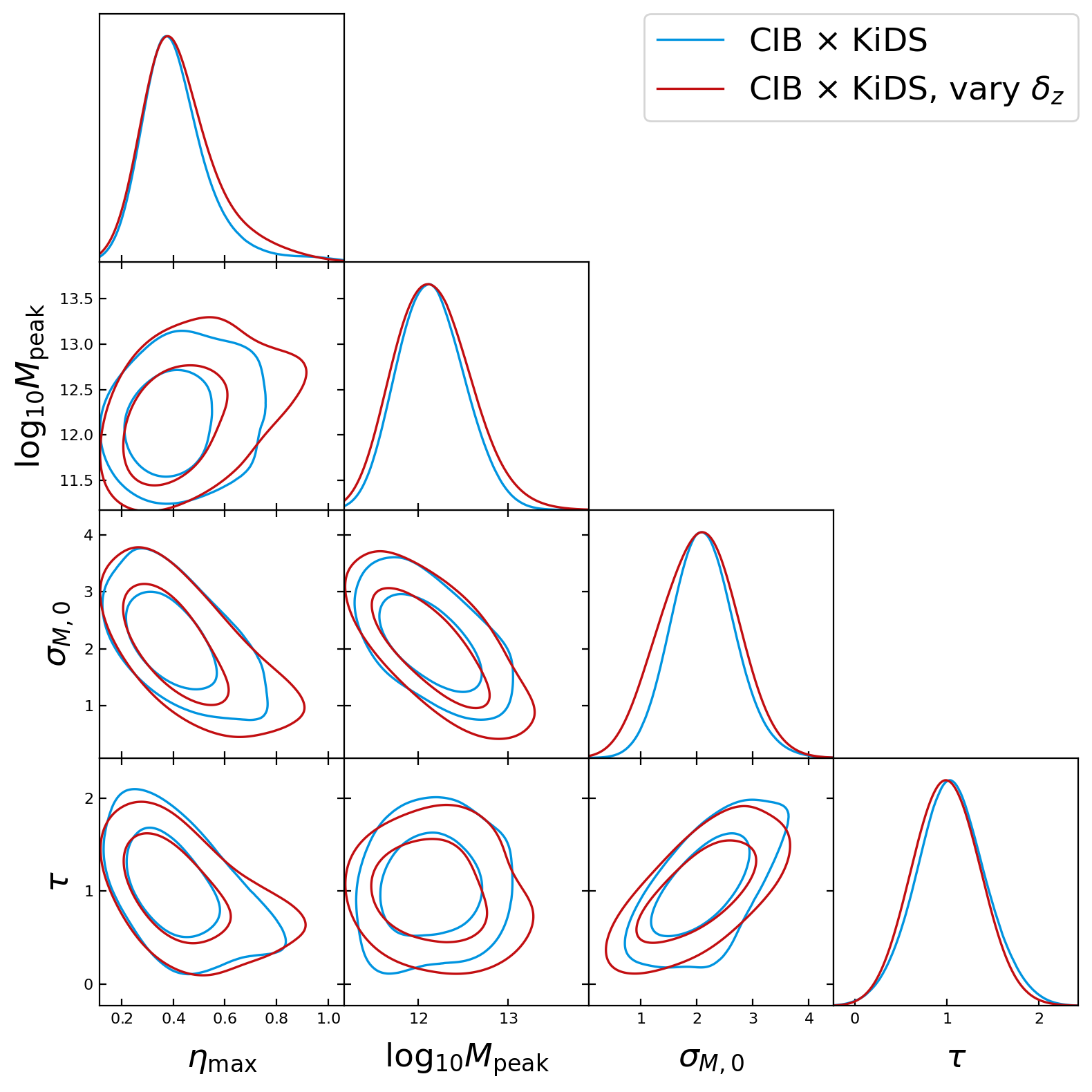}
    \caption{Posterior of SFR parameters from the \cco fit with fixed (blue) and freely varying (red) priors on the means of $\Phi_{\mathrm{g}}(z)$. The blue contour is our fiducial posterior.}
    \label{fig:params_dz}
\end{figure}
We also present the posterior of SFR parameters with varying $\delta_z$ in Fig. \ref{fig:params_dz}. The blue contour is our fiducial posterior, while the red contour is the posterior of the SFR parameters when allowing free variation of $\delta_z$ described in Sect.~\ref{sect:measurements}. Varying $\delta_z$ only slightly loosens the constraints on $\eta_{\mathrm{max}}$, while all other posteriors are largely unchanged. The posterior distributions of our  $\delta_{z,i}$ parameters follow the prior, demonstrating that there is no redshift-distribution self-calibration occurring in this analysis. Nonetheless, given the conservative choices made with our $\delta_{z,i}$ priors, we conclude that our constraints are robust to redshift distribution uncertainties in the galaxy sample. This result is largely expected, though, as CIB emission varies only slightly within the level of the $\delta_z$ uncertainties (see Fig. \ref{fig:dndz}).

\subsection{Constraint on galaxy bias}
\label{subsect:bg}
In this subsection we present the constraints on HOD parameters described in Sect. \ref{sect:model} and the derived galaxy bias. This is not the main scope of this paper, but {it is} nonetheless an interesting outcome of our study. Galaxy bias parameters are typically constrained using galaxy power spectra; however, this is challenging to perform with KiDS (and indeed with any full-depth galaxy survey) due to the complicated (artificial) {variable depth} between survey fields introduced by variable observing conditions. \citet{2021yanz} constrained the linear galaxy bias for KiDS using galaxy-CMB lensing cross-correlations, assuming a linear model. In this work, we derive the linear galaxy bias from the constrained HOD parameters.

The scale-dependent galaxy bias is defined as

\begin{equation}
    \left\langle b_{\mathrm{g}}(z,k)\right\rangle=\int \dr M  \frac{\dr n}{\dr M}b_{\mathrm{h}}(M,z)\left\langle \delta_{\mathrm{g}}(k,z|M)\right\rangle,
\end{equation}
where the galaxy density fluctuation $\delta_{\mathrm{g}}$ is the Fourier transform of Eq.~\eqref{eq:delta_g}. The linear galaxy bias is given by 

\begin{equation}
  \left\langle b^{\mathrm{lin}}_{\mathrm{g}}(z)\right\rangle \equiv \left\langle b_{\mathrm{g}}(z,k\rightarrow0)\right\rangle
.\end{equation}

\begin{figure}
    \centering
    \includegraphics[width=\columnwidth]{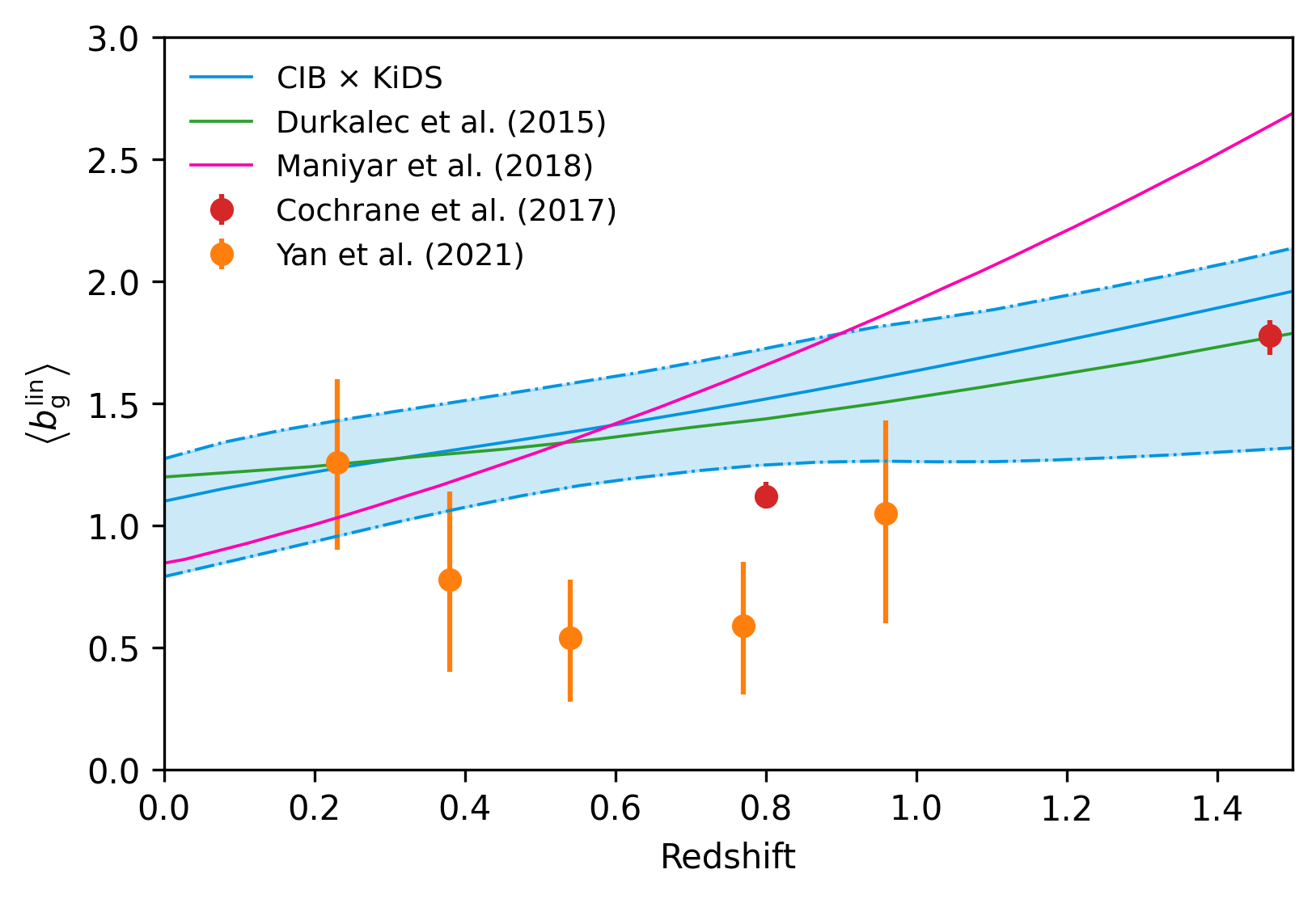}
    \caption{Linear galaxy bias constrained from CIB-galaxy cross-correlation. The solid blue line shows the best-fit $\left\langle b^{\mathrm{lin}}_{\mathrm{g}}(z)\right\rangle$, and the band with dotted-dashed boundary shows the upper and lower 1$\sigma$ errors. {We also over-plot results from previous studies.} }
    \label{fig:bg}
\end{figure}

The constrained HOD parameters are presented in Table \ref{tab:hod_fit}. Similar to the calculation of the best-fit SFRD and its uncertainty, we calculate the best-fit and 1$\sigma$ credible interval of $\left\langle b^{\mathrm{lin}}_{\mathrm{g}}(z)\right\rangle$ and present it in Fig \ref{fig:bg}. The resulting linear galaxy bias increases from $1.1_{-0.31}^{+0.17}$ at $z=0$ to $1.96_{-0.64}^{+0.18}$ at $z=1.5$. {We also over-plot constraints from previous studies on galaxy bias of star-forming galaxies. The magenta line shows the best-fit model from \citet[][]{maniyar_star_2018}; the green line shows the best-fit `star-forming' model from \citet{2015bg}; the red points are the derived galaxy bias of star-forming galaxies from \citet{2017MNRAS.469.2913C}.} We find good agreement between our result and these studies. The evolutionary trend of galaxy bias measured in this work is also in agreement with \citet{2007irbias}. It should be noted that the constraint of galaxy bias worsens at high redshift, as our galaxy sample is limited to $z<1.5$. 

{The galaxy bias constrained from CIB-galaxy cross-correlation is slightly higher than that constrained from galaxy-CMB lensing given by \citet{2021yanz} {(see the orange points)}. Galaxy bias from this work also shows a stronger evolution. These might be due to the fact that in \citet{2021yanz} all the galaxies in the KiDS gold sample contribute to the constraint, while the CIB-galaxy cross-correlation in this work is mainly sensitive {to KiDS galaxies that are active in star formation.} The fact that CIB cross-correlation gives higher galaxy bias means that these star-forming galaxies are (on average) more clustered than the galaxies {detected by optical survey}, especially at high redshift, which calls for further study. }

\begin{table}
\caption{Summary of the prior range, the mean, and 68\% confidence level of the HOD parameters.}
    \label{tab:hod_fit}
    \centering
    \begin{tabular} { l c c}
\toprule
 Parameter &  Prior &   68\% limits\\
\hline
{$\lgt M_{\mathrm{min,0}}        $} & [9, 14] & $10.7^{+1.3}_{-0.6}       $\\

{$\lgt M_{0,0}          $} & [9, 14] & $10.8^{+0.9}_{-1.6}      $\\

{$\lgt M_{1,0}          $} & [9, 14] & $12^{+1.6}_{-1.9}        $\\

{$\lgt M_{\mathrm{min,p}}       $} & [-5, 5] & $-1.3^{+1.3}_{-2.7}        $\\

{$\lgt M_{0,p}          $} & [-5, 5] & $0.7^{+3.8}_{-1.9}         $\\

{$\lgt M_{1,p}          $} & [-5, 5] & $1.0^{+3.7}_{-1.7}         $\\
\bottomrule
\end{tabular}
    
\end{table}

\section{Conclusions}
\label{sect:discussions}

In this work we measure the tomographic cross-correlation between the KiDS-1000 gold galaxy sample and the CIB maps made from \planck\ data. The motivation of this work is to provide a novel measurement of cosmic star formation history. We summarise our main conclusions in this section.

\begin{itemize}
    \item {The cross-correlation has a significance of 43$\sigma$, which is impressive given that the CIB signal is relatively low in the redshift range of our galaxy sample.}  Our CIB model yields an RCS value of 1.14 when fitting to the KiDS-CIB cross-correlation. Given the degrees-of-freedom, this corresponds to a PTE value of 0.13, meaning that the model fits the data well. 
    
    \item The constraints on the SFRD parameters from cross-correlation agree with those measured from external data, demonstrating that cross-correlation provides a novel, independent, and consistent probe of the cosmic star formation history. Moreover, this indicates that the halo model proposed by \citetalias{maniyar_simple_2021} is valid for both CIB cross-correlations and multi-wavelength studies of galaxies. 
    
    \item With our cross-correlation measurement, the maximum star formation efficiency is $\eta_{\mathrm{max}}=0.41^{+0.09}_{-0.14}$, in agreement with \citetalias{maniyar_simple_2021}. Our \cco-only measurement is unable to yield tight constraints on $M_{\mathrm{peak}}$, the halo mass that hosts the highest star formation efficiency, due to the galaxy sample being limited to $z<1.5$. A combination of cross-correlation and external SFRD measurements (the \ccosfrd constraints), however, tightens the constraint on $\log_{10}M_{\mathrm{peak}}$ to $\log_{10}M_{\mathrm{peak}}=12.42^{+0.35}_{-0.19}$, in agreement with previous studies within the 2$\sigma$ level (albeit with a preference for a slightly lower $M_{\mathrm{peak}}$ at posterior maximum). This may be due to the different data used in this study, which would imply  that measurements of $M_{\mathrm{peak}}$ are dependent on galaxy types. We leave an exploration of this for future work. Moreover, the best-fit $M_{\mathrm{peak}}$ from both \cco and \ccosfrd are in mild tension with \citetalias{maniyar_simple_2021}, which calls for further investigation. 
    
    \item We derived the SFRD from our posterior constraints over our various model parameters. The \cco constraint on the SFRD history is poor at high redshift because of sky coverage and depth limitations. Nevertheless, we note that in the redshift range probed by our sample of KiDS galaxies ($z<1.5$, corresponding to a lookback time of $\sim 10$ Gyr), cross-correlations give an SFRD that is consistent with previous galaxy-based multi-wavelength studies, at comparable constraining power. The \ccosfrd results tighten the constraint on the SFRD at all redshifts and are consistent with the SFRD from \cco and previous studies. We also compare the SFRD from this work with simulations and SAMs and find that our \cco constraints have the same trend as all the simulated SFRDs; however, our results are not sufficiently precise to invalidate any one model. Moreover, none of the simulations agrees with our \ccosfrd constraint at all the times, highlighting the complexity of the physical processes that underpin star formation activity in galaxies. 
    
    \item We also constrain the linear galaxy bias for KiDS galaxies that have significant IR emission. As for the SFRD, we can only constrain galaxy bias below $z\sim 1.5$, and with about 25\% precision. The constraint is limited by both the sky coverage and angular resolution of the CIB map. We also note that we model the redshift dependence of HOD parameters as a simple linear model with respect to the scale factor $a$, which could be improved for future studies. We derived the linear galaxy bias from the constrained HOD parameters, yielding an increasing galaxy bias greater than one. {The evolution of galaxy bias constrained in this work agrees with \citet{2015bg}, \citet{2017MNRAS.469.2913C}, and \citet{maniyar_star_2018}}. 
    
    \item For systematics, we took the colour correction of CIB flux and the effects of cosmic magnification into account. We also investigated the robustness of our results to the uncertainties in redshift distributions by allowing for shifts in the redshift distribution means. We find that this does not affect our constraints and conclude that our results are robust to the uncertainty in redshift distribution calibration. However, for future studies with higher signal-to-noise, this may become important.

\end{itemize}

\section{Discussions and future prospects}

{In this work we adopt the halo model for the CIB from \citetalias{maniyar_simple_2021}, with a minor modification such that the HOD model is consistent with \citet{Zheng_2005}. This model includes the information on the dust SED, star formation history, and galaxy abundance. Compared with other CIB models such as \citet{shang_improved_2012} and \citet{2020caoye}, it clearly formulates the redshift and halo mass dependence of SFR, which allows us to constrain the cosmic star formation history from cross-correlations between CIB and large-scale structure surveys.}

We make several assumptions to simplify the CIB halo model. For example, we assume that the SFR-IR luminosity connection can be described by a single Kennicutt parameter in the IR bands \citep[][]{kennicutt1998star}, with the assumption of a Chabrier IMF \citep{2003PASP..115..763C}. The SFR is modelled as the star formation efficiency times the baryonic accretion rate, which can be alternatively modelled by treating IR and UV components separately and introducing the  stellar-mass function \citep{bethermin_redshift_2013}. In addition, we have assumed that the SFR has the same mass dependence for central and satellite galaxies. In this work, we take the SED model from \citet{bethermin_evolution_2015}, which does not include the AGN contribution that could contaminate the mid-IR signal at $z>2$ \citep{bethermin_impact_2017}. This is beyond the scope of this paper, but future studies at these redshifts and frequencies should consider the influence of such assumptions. We do not discuss the thermodynamic properties of extragalactic dust, which are encoded in the SED. \citet{shang_improved_2012} provide an empirical model of the SED but do not model the SFR explicitly. Redshift-dependent SED changes and the evolution of the SFR are degenerate in the CIB signal, which might be resolved by introducing an additional tracer, or introducing the CIB at more frequencies (e.g. from the \textit{Herschel} and \textit{Spitzer} telescopes). We also note that the fit is worse at low redshift, which may indicate a limitation of our simplified model, or indicate the inaccurate measurement of our SED at low redshift. Finally, we have fixed all the cosmological parameters in this study. More sophisticated modelling is left for future studies that will use larger datasets.

The KiDS galaxy sample in this study has the advantage of extensive survey depth. Although our sample is not deep enough to directly constrain $M_{\mathrm{peak}}$, it yields a sensible measurement of the star formation history out to $z=1.5$. From Fig. \ref{fig:dndz}, we see that the redshift distribution of the KiDS galaxies is mostly in the {rising} regime of the CIB signal, which peaks at $z\sim 1.5$. For future galaxy surveys that will be able to reach $z\sim 3$, such as the Rubin Observatory  LSST and \euclid,  one would expect a more pronounced CIB-galaxy cross-correlation signal. In this context, we perform a forecast on the constraining power of the ongoing CFIS and the future LSST survey in Appendix \ref{sec:app1}. We find that the improvement of sky coverage makes CFIS yield a similar constraining power as our \ccosfrd constraint, while the more extensive LSST depth makes it possible to tightly constrain all the SFR parameters. The Dark Energy Survey has published its year-3 data, and CFIS will provide data with larger sky coverage in the near future, which will certainly help in clarifying the statistical significance of our \cco result without adding the external SFRD measurements. Furthermore, the LSST-CIB cross-correlation will be a promising tool, yielding enough constraining power to validate different SFR models and give us insight into the physics underpinning the cosmic star formation history.

Other prospective CIB studies include cross-correlating the CIB with galaxy samples as a function of brightness, colour, stellar mass, or morphology, as it is well known that star formation depends heavily on these properties \citep[as examples,][for brightness dependence; \citealt{2009mahajan} for colour dependence; \citealt{2018PASJ...70....4K} for stellar mass dependence; and \citealt{2013tojeiro} for colour-morphology combination dependence]{2012yajima}. In addition, dust properties should also depend on these properties \citep{2009wolf, noll2005dust}. A CIB cross-correlation with different types of galaxies could therefore serve as a new independent probe of these properties.

Central and satellite galaxies are located in different physical environments, resulting in different SFRs. Specifically, AGN feedback and quenching are two interesting processes that can effect a galaxy's SFR, and they are found to be different in central and satellite galaxies \citep[][]{2018quenching}. In this work we do not separately study the SFR of central and satellite galaxies because the total SFR is dominated by central galaxies with higher halo mass, and the SFR of satellite galaxies is not well constrained. This may be improved with future surveys. Once more significant cross-correlation measurements are available, we will be able to study quenching and AGN activities in central and satellite galaxies by adding more parameters that describe these effects to the SFR model.

{The constraints on HOD parameters suggest an increasing galaxy bias through redshift. At high redshift, the high galaxy bias might be due to the triggering of star formation by mergers in dense environments at $z\sim 1$ \citep{2007ApJ...656..139W}. At low redshift, however, star formation is quenched through gas stripping in dense regions \citep{2005galaxybias}, leading to a lower overall bias. In conclusion, a comparison between the galaxy bias of normal galaxies and star-forming galaxies indicates the evolution of the link between star formation and environments. It should be noted that the constraining power of HOD parameters in this study is weak because of the limitation of both sky coverage and depth. Moreover, the linear formalism of HOD parameters is a simplified empirical model. Future studies with improved sky coverage and depth should improve the constraints of $\left\langle b^{\mathrm{lin}}_{\mathrm{g}}(z)\right\rangle$.}

In summary, the CIB is a gold mine that encodes information about the star formation history of the Universe, extragalactic dust properties, and galaxy abundance. This work validates the CIB-galaxy cross-correlation method as a valuable tool for understanding the cosmic star formation history. The success of our measurement here, despite the limitations discussed above, provides an exceedingly positive outlook for future analyses of the CIB-galaxy cross-correlation. Larger, deeper datasets, coupled with more complex sample subdivisions, will allow us to leverage CIB cross-correlations to better understand the growth and evolution of galaxies in our Universe.

\begin{acknowledgements}
We would like to thank Yunhao Zhang, Dr. Gary Hinshaw, and Dr. Joachim Harnois-Déraps for useful discussions. We thank the Planck Collaboration for the all-sky data available at \href{https://www.cosmos.esa.int/web/planck/pla}{https://www.cosmos.esa.int/web/planck/pla} and Dr. Daniel Lenz for making the CIB maps. \\
Based on observations made with ESO Telescopes at the La Silla Paranal Observatory under programme IDs 177.A-3016, 177.A-3017, 177.A-3018 and 179.A-2004, and on data products produced by the KiDS consortium. The KiDS production team acknowledges support from: Deutsche Forschungsgemeinschaft, ERC, NOVA and NWO-M grants; Target; the University of Padova, and the University Federico II (Naples).\\
ZY acknowledges support from the Max Planck Society and the Alexander von Humboldt Foundation in the framework of the Max Planck-Humboldt Research Award endowed by the Federal Ministry of Education and Research (Germany). LVW and SG acknowledge acknowledge the support support by the University of British Columbia, Canada's NSERC, and CIFAR. MB is supported by the Polish National Science Center through grants no. 2020/38/E/ST9/00395, 2018/30/E/ST9/00698, 2018/31/G/ST9/03388 and 2020/39/B/ST9/03494, and by the Polish Ministry of Science and Higher Education through grant DIR/WK/2018/12. H. Hildebrandt is supported by a Heisenberg grant of the Deutsche Forschungsgemeinschaft (Hi 1495/5-1) and, along with A.H. Wright, by an ERC Consolidator Grant (No. 770935). TT acknowledges support from the Leverhulme Trust\par
\\
The data in this paper is analysed with open-source python packages \textsc{numpy} \citep{harris2020array}, \textsc{scipy} \citep{2020SciPy-NMeth}, \textsc{astropy} \citep{astropy:2018}, \textsc{matplotlib} \citep{Hunter:2007}, \textsc{healpy} \citep{Zonca2019}, \textsc{NaMaster} \citep{2019namaster}, \textsc{CCL} \citep{Chisari_2019}, \textsc{emcee} \citep{Foreman_Mackey_2013}, and \textsc{GetDist} \citep{Lewis:2019xzd}.
We also use \textsc{WebPlotDigitizer} \citep{Rohatgi2020} to digitize some external data from plots in the literature.
\\
{{\it Author contributions:} All authors contributed to the development and writing of this paper. The authorship list is given in three groups: the lead authors (ZY \& LvW) followed by two alphabetical groups.  The first alphabetical group includes those who are key contributors to both the scientific analysis and the data products. The second group covers those who have either made a significant contribution to the data products, or to the scientific analysis.
 }\\
{{\it Data availability:} The cross-correlation data and the MCMC chain of this work will be shared on reasonable request to the corresponding author.}
\end{acknowledgements}

\bibliographystyle{aa}
\bibliography{main}

\begin{appendix}

\section{The jackknife covariance matrix}
\label{app:jackcov}
\begin{figure*}
    \centering
    \includegraphics[width=\textwidth]{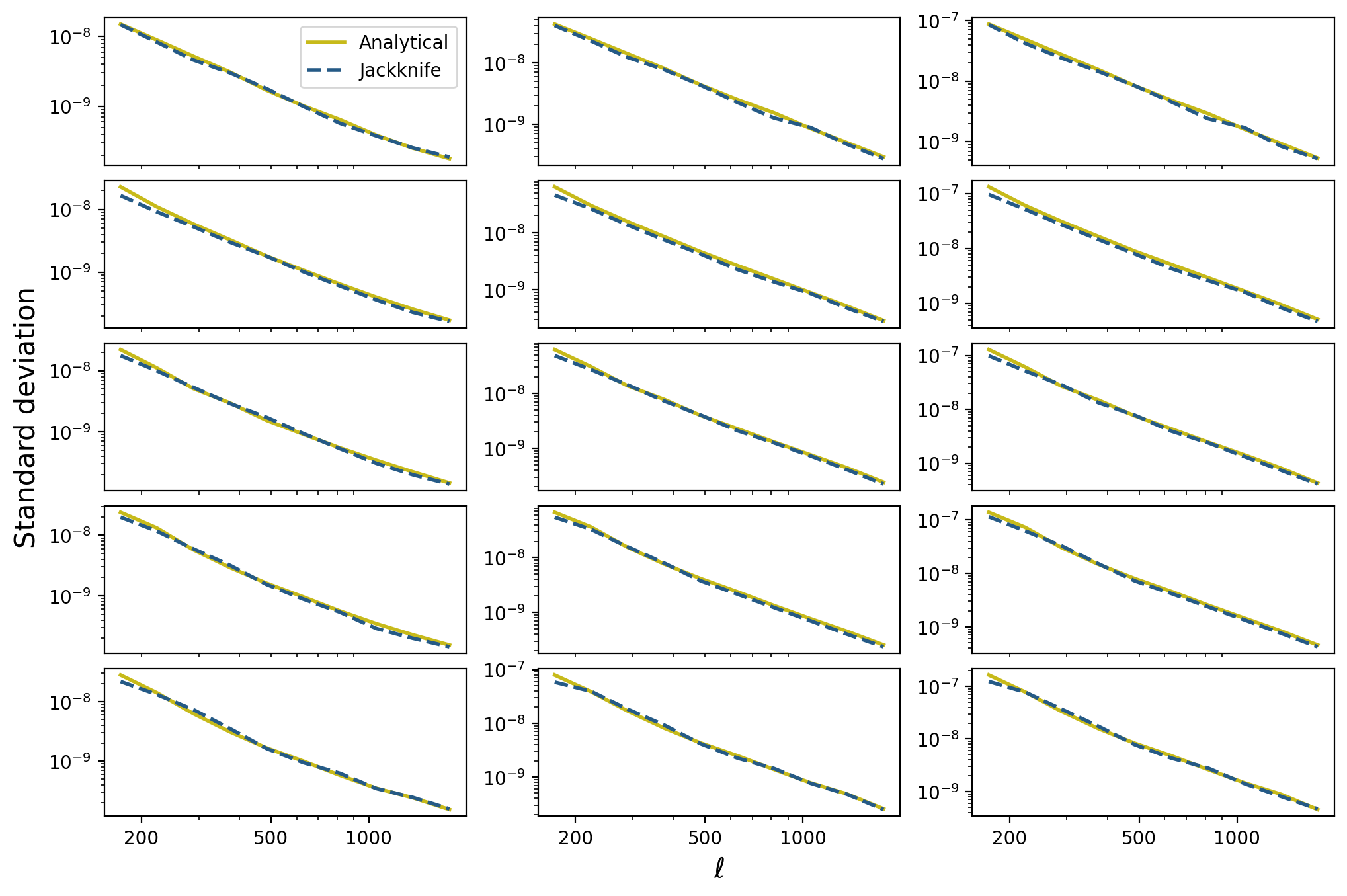}
    \caption{Comparison between the standard deviation of $C^{\nu\mathrm{g}}_{\ell}$ obtained from the diagonal terms of the analytical (the beige lines) and jackknife (the dashed dark blue lines) covariance matrices. Each panel corresponds to one cross-correlation signal. The three columns represent three CIB channels in increasing order from left to right; the five rows represent five KiDS tomographic bins in increasing order from top to bottom.}
    \label{fig:std_compare}
\end{figure*}

An alternative method to estimate the covariance matrix is to use jackknife resampling, as is used in \citet{2021yanz}. The idea is to resample the signal by removing a small patch of the sky and calculate the cross-correlation from the remaining sky. We remove one patch at a time with replacement to get several cross-correlation signals. The covariance matrix is estimated by calculating the covariance matrix from these resampled cross-correlations. Except for assuming that all the patches are independent, this method does not depend on any other prior knowledge of the signal and is in principal able to take all the contributions into the account. However, there are two drawbacks of this method. Firstly, the KiDS footprint is small, so we can only have a small size of the jackknife sample. If we increase the sample by using small jackknife patches, we will lose accuracy in covariance corresponding to large-scale modes. This problem is more severe in this study since we only use nearly half of the KiDS field. Secondly, removing patches from the sky changes the shape of the footprint, which will change the coupling matrix of cross-correlations, thus biasing the covariance matrix. Therefore, we only use jackknife covariance as a consistency check of our analytical covariance matrix in this work. We generate 220 jackknife samples by removing \textsc{Healpix} pixels with \textsc{Nside=32}, which correspond to a size of 1.8 degree. A comparison between standard deviations calculated from the analytical and jackknife covariance matrices is shown in Fig. \ref{fig:std_compare} from which we see that both covariance matrices agree well, with small deviations in large scales. We note that this scale ($\ell\sim 100$) is close to the size of removed pixels, so this deviation might be due to the poor sampling of large scales in our jackknife scheme.

\section{Forecast for future surveys}
\label{sec:app1}

As has been discussed in the main text, the SFRD constraint from CIB-galaxy cross-correlation in this study is mainly limited by sky coverage and survey depth. Fortunately, future surveys will have much wider coverage and higher redshift depth. In this section we forecast the constraining power on the SFRD parameters from CIB-galaxy cross-correlations with galaxies from future surveys. We adopt Fisher forecast by constructing the Fisher matrix:

\begin{equation}
    \mathrm{F}_{\alpha \beta} =\frac{\partial \boldsymbol{C}^{\mathrm{T}}}{\partial q_{\alpha} }\mathrm{Cov}^{-1}\frac{\partial \boldsymbol{C}}{\partial q_{\beta}},
\end{equation}
where $\boldsymbol{C}(\boldsymbol{q})$ is again the model cross-correlation with parameters $\boldsymbol{q}$. The covariance matrix is calculated the same way as discussed in Sect. \ref{subsect:covmat}. Several factors in the covariance matrix are survey-specified, namely sky coverage fraction, shot noises, and galaxy redshift distributions. The sky coverage fraction and galaxy redshift distributions can generally be found in survey proposal papers (which is to be introduced below). Shot noise (needed in Gaussian covariance matrix) for galaxy-galaxy power spectra is calculated via

\begin{equation}
    \mathrm{SN}^{\mathrm{gg}}_{\ell} = 1/\overline{N},
\end{equation}
where $\overline{N}$ is the mean number of galaxies per steradian, which can also be found in the survey proposal, and we estimate the galaxy density in each tomographic bin by weighting the total galaxy density by redshift distributions of each bin. The CIB-CIB shot noise is taken from the constraints given by \citetalias{maniyar_simple_2021}. CIB-galaxy shot noise are estimated as the best-fit shot noise corresponding to galaxy sample in the similar redshift bin from this work. For CIB-galaxy shot noise, one would expect that it should also be positively correlated with $1/\overline{N}$. For future surveys, this should be lower. However, without precise knowledge of this dependence, we use the CIB-galaxy shot noise for the KiDS survey as a conservative estimation.

We consider two ongoing and future galaxy surveys. The first is the LSST based at Vera Rubin Observatory \citep{lsstsciencecollaboration2009lsst}.\ It is one of the major fourth-generation cosmological surveys that cover about 20000 deg$^2$. The overlap of LSST and the L19 CIB map is about 8\% of the sky. The galaxy density is estimated to be 55.5 $\mathrm{arcmin}^{-2}$. We adopt the overall redshift distribution and tomographic binning model from \citet{lsstsciencecollaboration2009lsst}. We take ten tomographic bins from $z=0$ to $z=3$.

The second is the CFIS \citep{2017cfis}, an ongoing cosmological survey based at the Canada-France-Hawaii Telescope. When completed, it will yield a sky coverage of 3500 deg$^2$, overlapping with the L19 CIB map about 4\% of the sky. The average galaxy density is about 10 $\mathrm{arcmin}^{-2}$. The total redshift distribution is given by \citet{SpitzerIsaac2022}. We adopt the same tomographic binning model as LSST with five tomographic bins from $z=0$ to $z=1.5$.

\begin{figure*}
    \centering
    \includegraphics[width=0.8\textwidth]{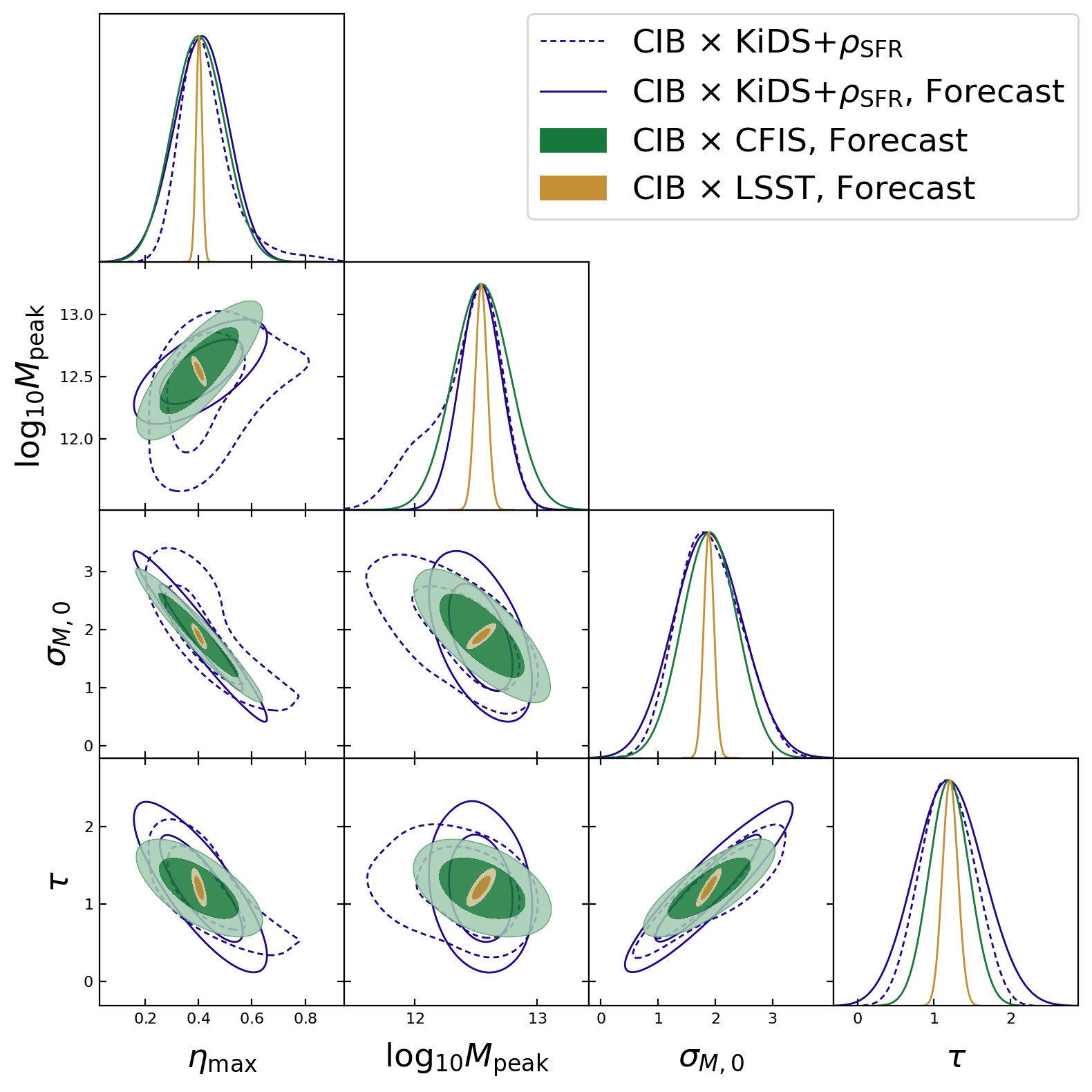}
    \caption{Fisher forecast for CIB-galaxy cross-correlation with LSST (green) and CFIS (dark yellow), as well as the `forecast' for the \ccosfrd measurement in this work (solid purple contour). The \ccosfrd constraint is also shown as a dashed purple contour. Contours are the 1 and 2$\sigma$ levels.}
    \label{fig:fisher}
\end{figure*}

We show the Fisher forecast contours in Fig. \ref{fig:fisher}. The green and the dark yellow contours correspond to CIB-galaxy cross-correlations with future CFIS and LSST galaxy catalogues, respectively. The contours are 1$\sigma$ and 2$\sigma$ levels of confidence. We also plot the `forecast' for our \ccosfrd measurement with solid purple contours and the constraints from this work with dashed purple contours. We see that LSST yields much better constraints of all the SFR parameters while CFIS yields comparable constraints as \ccosfrd (therefore both better than \cco constraint).

We note that the maximum star formation efficiency $\eta_{\mathrm{max}}$ acts as an overall normalisation parameter, so its constraint is mainly related to sky coverage. The rest three parameters control the redshift dependence of SFRD (and thus the CIB emission). Therefore, a deeper survey would give a better constrain on them, especially $M_{\mathrm{peak}}$ (see Fig. \ref{fig:sfrd_varym}). The constraints on these three parameters are not significantly improved with CFIS because CFIS has a similar depth as KiDS. LSST goes much deeper, so it can provide much better constraints.

The noise levels and redshift distributions in this forecast study are over-simplified. Moreover, the difference between the solid and the dashed purple contours shows that the posteriors are not Gaussian; therefore, the Fisher forecast might be inaccurate to predict the posterior. However, the parameter errors from the solid purple contours do not differ much from the measured errors, so the forecasts are still useful. The main information that we learn from the forecast is that sky coverage and survey depth significantly affects the SFR constraints from CIB tomography. The ongoing CFIS survey is promising to yield similar constraining power as \ccosfrd, and LSST can achieve much tighter constraints. In this forecast study, we only consider the L19 CIB map with a relatively low angular resolution. LSST overlaps with \textit{Herschel}/SPIRE \citep{2013ApJ...772...77V}, which {detects IR galaxies with} a higher resolution at a level of $\sim 10$ arcsec. This makes it possible to study galaxy clustering on smaller scales. Next-generation CMB surveys (CMB-S4, for example) will also achieve higher angular resolution. In addition, Roman Space Telescope and James Webb Space Telescope will thoroughly explore the near-IR sky. \euclid\, will also probe IR-selected galaxies at high redshifts, which should be useful for CIB studies. When combined with other probes, these new data will provide us with better insight into the cosmic star formation history and the complicated physics behind it.

\end{appendix}

\end{document}